\documentclass[a4paper,12pt]{article}
\usepackage{jheppub} 
\usepackage{lineno}
\usepackage[]{babel}
\usepackage[normalem]{ulem}
\makeatletter
\def\@fpheader{\relax}

\newcommand{\gammah}{\alpha_h}
\newcommand{\betah}{\beta_h}
\newcommand{\alphah}{\gamma_h}
\newcommand{\alphak}{\beta_k}
\newcommand{\betak}{\alpha_k}
\newcommand{\gammak}{\gamma_k}
\newcommand{\AB}[1]{{\color{black}{#1}}}
 
 \newcommand{\AM}[1]{{\color{black}{#1}}}

\title{Nambu-Goto equation from three-dimensional gravity }

\author[a,b]{Avik Banerjee,}
\author[c,d]{Ayan Mukhopadhyay,} \author[b]{and Giuseppe Policastro}
\affiliation[a]{Crete Center for Theoretical Physics, Institute for Theoretical and
Computational Physics, Department of Physics, University of Crete, Heraklion, Greece}
\affiliation[b]{Laboratoire de Physique de l'\'{E}cole Normale Supérieure, ENS, Universit\'{e} PSL, CNRS, Sorbonne Universit\'{e}, Universit\'{e} de Paris, F-75005 Paris, France}
\affiliation[c]{Instituto de F\'{\i}sica, Pontificia Universidad Cat\'{o}lica de Valpara\'{\i}so, Avenida Universidad 330, Valpara\'{\i}so, Chile}
\affiliation[d]{Department of Physics, Indian Institute of Technology Madras, Chennai 600036, India}

\abstract{We demonstrate that the solutions of three-dimensional gravity obtained by gluing two copies of a spacetime across a junction constituted of a tensile string are in one-to-one correspondence with the solutions of the Nambu-Goto equation in the same spacetime up to a finite number of rigid deformations {related to worldsheet and spacetime isometries.} The non-linear Nambu-Goto equation satisfied by the average of the embedding coordinates of the junction emerges directly from the junction conditions along with the rigid deformations and corrections due to the tension. Therefore, the equivalence principle generalizes non-trivially to the string. Our results are valid both in three-dimensional flat and AdS spacetimes. In the context of AdS$_3$/CFT$_2$ correspondence, our setup could be used to describe a class of interfaces in the conformal field theory featuring relative time reparametrization at the interface which encodes the solution of the Nambu-Goto equation corresponding to the bulk junction.}

\begin{document}
\maketitle
\flushbottom

\section{Introduction}\label{Sec:Intro}
The equivalence principle follows from the consistency of Einstein's theory of gravity. The conservation of the stress tensor of the point particle, which is necessitated by the Bianchi identity, implies that the particle should move along a geodesic. The Nambu-Goto equation generalizes the geodesic equation by implying that the string's motion should extremize its worldsheet area. Could the Nambu-Goto equation also follow from gravity? 

We examine this question in pure three-dimensional gravity in flat and AdS spaces. We consider solutions which are obtained by gluing two copies of a spacetime $\mathcal{M}$ across a junction constituted of a  string with a finite tension. We find that such solutions of pure gravity are in one-to-one correspondence with the solutions of the Nambu-Goto equation for the motion of the string in $\mathcal{M}$ up to corrections due to the finite tension and a \textit{finite} number of \textit{possible} rigid deformations. By rigid deformations, we mean additional contributions that arise due to the worldsheet isometries or are related to displacements of a hypersurface {via isometries of the embedding spacetime} which preserve its extrinsic curvature.\footnote{We discuss possible physical interpretations of the rigid deformations especially in Sec. \ref{Sec:Disc}.} In particular, we show that the Nambu-Goto equation is recovered in the tensionless limit in absence of the rigid deformations.

The equivalence principle thus generalizes non-trivially to the string since the inclusion of the backreaction on the three-dimensional spacetime due to the tension does not preserve the extremality of the worldsheet area for the motion of the string.  However, when the string tension and rigid deformations vanish, a solution of the Nambu-Goto equation for the string in $\mathcal{M}$ is obtained by taking the \textit{average} of the embedding  coordinates of the junction in the two copies of $\mathcal{M}$.  We will demonstrate these claims by employing a perturbative expansion around trivial solutions of the junction conditions. It should be possible to obtain a non-perturbative proof, but we will leave this for future work.

Our results lead to a fresh perspective on Einstein's dilemma of whether to consider the right hand side of his eponymous equations of gravity as \textit{ugly}. The usual formulation of string theory prioritizes the right hand side. It geometrizes and unifies matter as quantum vibrations of a fundamental string, and also makes semi-classical gravitational spacetimes emerge from the worldsheet. However, our results imply that the classical motion of the fundamental string is itself a consequence of spacetime dynamics in three dimensions. Therefore, it is natural to ask whether the spectrum and quantum dynamics of the (first quantized) fundamental string \cite{Polchinski:1998rq,Polchinski:1998rr,Maldacena:2000hw,Maldacena:2000kv,Gaberdiel:2018rqv} can also be part of quantized \textit{pure} three-dimensional gravity in which junctions with a finite tension are included.\footnote{Since we are discussing three-dimensional gravity, we cannot hope to recover the full spectrum of the fundamental string in critical number of dimensions.}

Furthermore, the holographic correspondence states that the classical gravitational dynamics of pure AdS$_3$ can be described by a universal sector of two-dimensional conformal field theories with large central charges (see \cite{Kraus:2006wn} for a review with discussions on applications to black hole physics). It is then also natural to ask if the Nambu-Goto equation can also emerge from dynamical interfaces in conformal field theories with large central charges implying that the fundamental string can be found directly in the dual field theory.\footnote{Remarkably, a precise version of AdS$_3$/CFT$_2$ correspondence has been derived in \cite{Eberhardt:2018ouy,Eberhardt:2019ywk} in which the quantum gravity in AdS$_3$ is described by a superstring theory. In this case, the spacetime has extra compact directions (S$^3 \times$ T$^4$) and also one unit of NS-NS flux, and the dual CFT can be written in terms of free fields. Here, we are in the limit in which the gravitational dynamics is classical and described by Einstein's equations so that the dual CFT has a large central charge and is also strongly interacting. Quantization could require the presence of extra dimensions for consistency. However, the three-dimensional classical solutions with junctions described here should be dual to interfaces between states in the universal sector involving the Virasoro identity block only.} As discussed later, our solutions with junctions in anti-de Sitter space can indeed have holographic interpretations as a class of interfaces in the dual conformal field theory where there is relative time-reparametrization at the interface encoding the Nambu-Goto solution corresponding to the bulk junction. In this way we generalize previous works that used backreacted strings with tension in AdS  as bottom-up models of conformal defects and boundaries, see \cite{Karch:2000ct,DeWolfe:2001pq,Takayanagi:2011zk, Bachas:2020yxv,Bachas:2021tnp,Liu:2024oxg}. 

In the rest of the paper, we proceed by first discussing the setup of our calculations and then giving a precise statement of our results in Section \ref{Sec:Setup}. Subsequently, we demonstrate the one-to-one correspondence between the solutions of the Nambu-Goto equation and the gravitational solutions with junctions carrying finite tension in three-dimensional flat space and anti-de Sitter space in Sections \ref{Sec:FS} and \ref{Sec:AdS}, respectively. Finally, we conclude with discussions on the implications of our results in Section \ref{Sec:Disc}.

\section{Setup and statement of results}\label{Sec:Setup}

\subsection{The question and why $D=3$ is special}
Consider a junction $\Sigma$ constituted by a  co-dimension one brane with tension $T_0$ embedded in a $D$-dimensional manifold $\mathcal{M}$. The full manifold $\mathcal{M}$ involves the union of two $D$-dimensional spacetimes $\mathcal{M}_L$ and $\mathcal{M}_R$ which are glued at the brane junction $\Sigma$. The Einstein-Hilbert action with a cosmological constant with the brane source is
\begin{align}\label{Eq:Action}
    S = \frac{1}{16\pi G_N}\int_{\mathcal{M}}{\rm d}^D x \sqrt{-g} (R+ 2 \Lambda)
    +  T_0\int_\Sigma{\rm d}^{D-1} y \sqrt{-\gamma}
    + {\rm GHY}\,\, {\rm terms},
\end{align}
where $g$ is the bulk metric, $\gamma$ is the induced metric on $\Sigma$ and GHY terms stand for the Gibbons-Hawking-York boundary terms.\footnote{There are two GHY terms associated to $\Sigma$ as it is the part of the boundary of both $\mathcal{M}_L$ and $\mathcal{M}_R$. If there are additional boundaries, then there are more GHY terms. However, the latter will not be relevant for the current discussion.}  It is to be noted that the bulk metric $g$ is the only physical field in this action. The embedding of $\Sigma$  should be determined from the \textit{junction conditions} obtained from extremizing this action with respect to variations of the bulk metric at the junction, and not from separate equations of motion. Therefore, the limit $T_0 \rightarrow 0$ need not correspond to the probe brane in an Einstein manifold. 

The natural question is whether we can identify a generic solution of the gravitational theory described by the action \eqref{Eq:Action} with solutions of the worldvolume extremization equations of a co-dimension one brane in $\mathcal{M}$. We can readily argue this is not possible for $D \geq 4$, and that the case $D=3$ is special, as follows. 

Instead of using continuous bulk coordinates in which the normal and the tangents of $\Sigma$ are continuous across the junction, it is convenient for our purposes to adopt the methodology of \cite{Bachas:2020yxv} (but without restricting to only linear perturbations about a simple solution and assuming any specific boundary conditions). Following \cite{Bachas:2020yxv}, we adopt coordinates $(t_L, x_L, \vec{z}_L)$ for $\mathcal{M}_L$ and $(t_R, x_R, \vec{z}_R)$ for $\mathcal{M}_R$, where $t_{L,R}$ are the time coordinates, while $x_{L,R}$ and the $\vec{z}_{L,R}$ with $D-2$ components are the $D-1$ spatial coordinates in the respective halves. The hypersurface $\Sigma$ has two images in $\mathcal{M}_L$ and $\mathcal{M}_R$, which are $\Sigma_L$ and $\Sigma_R$ respectively, and should be glued via solving the junction conditions. The hypersurface $\Sigma_L$ in $\mathcal{M}_R$ is given by
\begin{equation}
  x_L = f_L(t_L, \vec{z}_L),
\end{equation}
Similarly, the hypersurface $\Sigma_R$ in $\mathcal{M}_R$ is given by
\begin{equation}
  x_R = f_R(t_R, \vec{z}_R).
\end{equation}
Furthermore, $\mathcal{M}_L$ is the set of points $x_L \leq f_L(t_L, \vec{z}_L)$ while $\mathcal{M}_R$ is the set of points $x_R \geq f_R(t_R, \vec{z}_R)$.

A point $P_L$ on $\Sigma_L$ is identified with another point $P_R$ on $\Sigma_R$ if these points carry the same brane coordinate labels $(\tau, \vec{\sigma})$. However, we also need to fix the diffeomorphisms on the brane. It is convenient to fix the diffeomorphisms on the brane via these gauge conditions:
\begin{eqnarray}
  \tau = \frac{1}{2}(t_L(P_L) + t_R(P_R)) , \quad \vec{\sigma} = \frac{1}{2}(\vec{z}_L(P_L) + \vec{z}_R(P_R)).  
\end{eqnarray}
Therefore, the parametric forms of $\Sigma_L$ and $\Sigma_R$ are
\begin{align}
    &t_L(\tau, \vec{\sigma}) = \tau - \tau_a(\tau, \vec{\sigma}), \quad \vec{z}_L(\tau, \vec{\sigma}) = \vec{\sigma} - \vec{\sigma}_a(\tau, \vec{\sigma}), \quad x_L(\tau, \vec{\sigma}) = f_L\left(t_L(\tau, \vec{\sigma}), \vec{z}_L(\tau, \vec{\sigma})\right),\nonumber\\
    &t_R(\tau, \vec{\sigma}) = \tau + \tau_a(\tau, \vec{\sigma}), \quad \vec{z}_R(\tau, \vec{\sigma}) = \vec{\sigma} + \vec{\sigma}_a(\tau, \vec{\sigma}), \quad x_R(\tau, \vec{\sigma}) = f_R\left(t_R(\tau, \vec{\sigma}), \vec{z}_R(\tau, \vec{\sigma})\right),\label{Eq:Sigma12a}
\end{align}
where $\tau_a= (1/2)(t_R - t_L)$ and $\vec{\sigma}_a =(1/2)(\vec{z}_R - \vec{z}_L)$ are the relative time and spatial shifts which are not fixed by the choice of gauge and should be determined via the junction conditions. In total, the junction conditions need to determine $D+1$ functions of $\tau$ and $\vec{\sigma}$, namely the $D-1$ relative coordinates $\tau_a$ and $\vec{\sigma}_a$, and $f_L$ and $f_R$ which give the images of the embedding of the brane. Solving the $D+1$ variables we can determine the geometric embedding of the brane in the bulk spacetime completely.

Explicitly by extremizing the action \eqref{Eq:Action}, we obtain Einstein's equations
\begin{equation}\label{Eq:EH}
    R_{MN} - \frac{1}{2} R g_{MN} + \Lambda g_{MN} =0,
\end{equation}
away from $\Sigma$, i.e. away from $\Sigma_{L}$ and  $\Sigma_{R}$ in $\mathcal{M}_L$ and $\mathcal{M}_R$, respectively. Above, $M$ and $N$ stand for the $D$ bulk indices. The extremization on the junction gives \cite{Israel:1966rt}
\begin{eqnarray}
    \gamma_{\mu\nu}\vert_{\Sigma_R}- \gamma_{\mu\nu}\vert_{\Sigma_L} &=& 0,\label{Eq:BJC1}\\
    (K_{\mu\nu}- K\gamma_{\mu\nu})_{\Sigma_R} - (K_{\mu\nu}- K\gamma_{\mu\nu})_{\Sigma_L} &=& 8\pi G_N T_0 \gamma_{\mu\nu},\label{Eq:BJC2},
\end{eqnarray}
where $\mu$ and $\nu$ are the $D-1$ brane indices, and $K_{\mu\nu}$ is the (respective hypersurface's) extrinsic curvature. The above are called the Israel junction conditions \cite{Israel:1966rt}. The combination $K_{\mu\nu} - K \gamma_{\mu\nu}$ is also called the Brown-York stress tensor \cite{Brown:1992br}.

The first set of junction conditions \eqref{Eq:BJC1} involving the continuity of the induced metric give $D(D-1)/2$ equations. Although the second set of junction conditions \eqref{Eq:BJC2} involving the discontinuity of the Brown-York stress tensor give $D(D-1)/2$ equations as well, $D-1$ of them are redundant as the Brown-York stress tensor of any hypersurface is identically conserved (in the background given by the induced metric) on an Einstein manifold by virtue of the Gauss-Codazzi equations. Therefore, the junction conditions give in total $$D(D-1) - (D-1) = (D-1)^2$$equations. However, we have already seen that the embedding of the junction is determined by $D+1$ variables. We readily observe that for $D\geq3$, $$(D-1)^2 \geq D+1,$$with this inequality saturated only when $D=3$. Therefore, for $D\geq 4$, there are more equations that are obtained from the junction conditions than the number of variables which determine the geometric embedding of the brane. This argument implies that we cannot recover \textit{generic} solutions of the worldvolume extremization equations of the brane from the junction conditions in the tensionless limit for $D\geq 4$.

The case $D=3$ is special as the junction conditions give 4 equations that can fully determine the 4 variables which in turn completely specify the geometric embedding of the brane. In this paper, we will show that {the naive counting argument is indeed correct and we} can identify any solution of the gravitational equations as the backreaction of a string which satisfies the Nambu-Goto equations in the probe limit ($T_0 \rightarrow 0$) up to a finite number of rigid deformation parameters in this case. We will describe our setup and summarize our results precisely in the following subsection.\footnote{The case $D=2$ is also special.  It can be shown that in this case, the general solutions involve gluing two copies of flat spaces ($\Lambda =0$), or two AdS$_2$ ($\Lambda <0$), or two dS$_2$ ($\Lambda >0$) spacetimes with identical metrics, but with an isometry performed on one of the two copies. The junction follows the geodesic equation and this geodesic is invariant under a one parameter family of isometries. The isometry which is operated on one of the two bulk copies of flat space, AdS$_2$ or dS$_2$ is within this one parameter family, and is determined by the conserved charge of the geodesic. (For every isometry, a geodesic has a conserved charge.)}

\subsection{Setup and results}

In this paper, we study the solutions of the action \eqref{Eq:Action} in $D=3$. However we also further specialize to the case of gluing two \textit{identical} copies of an Einstein spacetime satisfying the equations \eqref{Eq:EH} for the simplicity of understanding how the Nambu-Goto equation arises from the junction conditions. For the sake of clarity, we describe our setup from scratch instead of falling back upon the general discussion in the previous subsection although many aspects of the discussion are going to be repetitive. 

\begin{figure}[h!]
    \centering
    \includegraphics[scale=0.20]{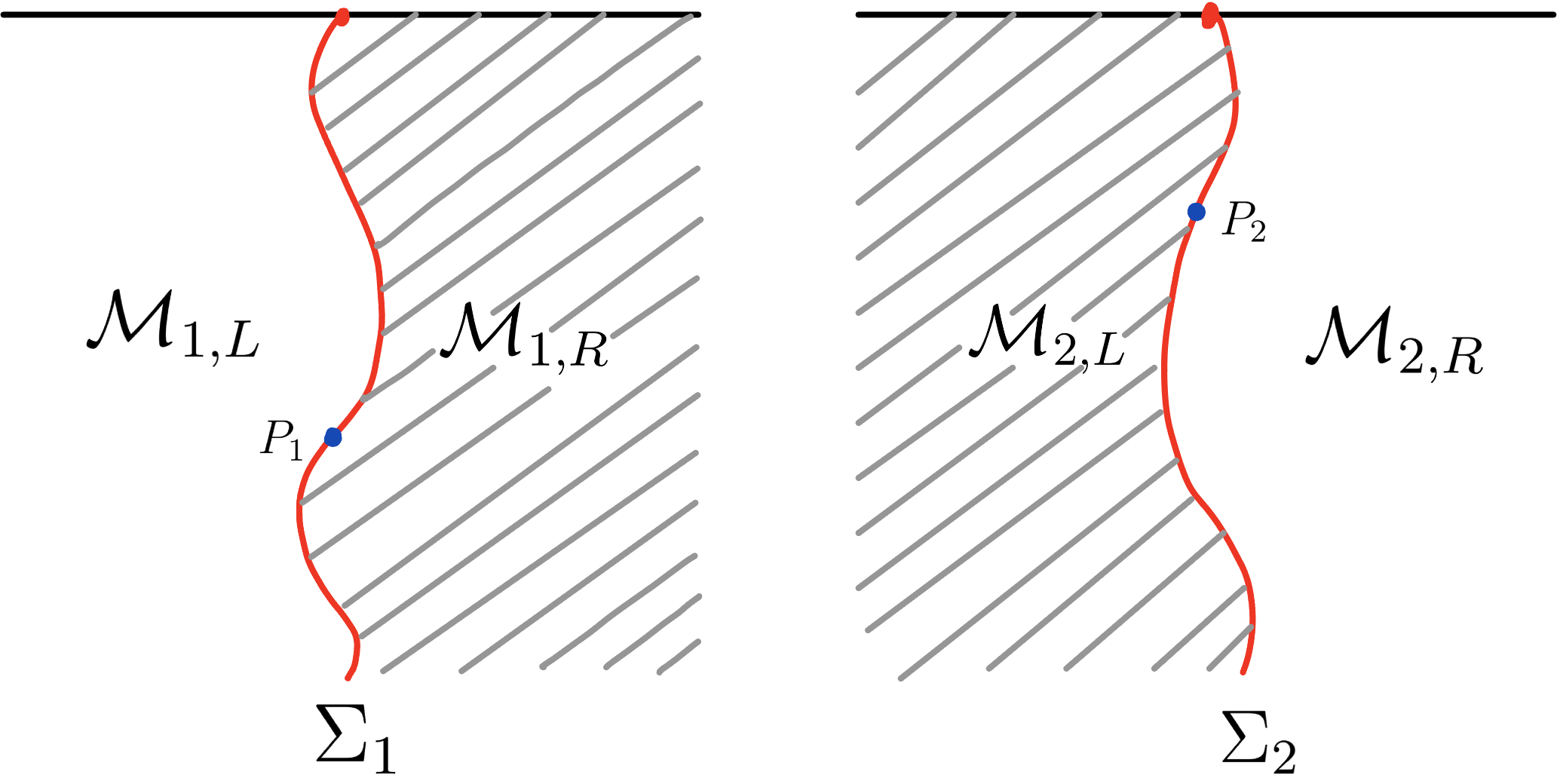}
    \caption{Illustration of the setup in AdS. We consider two copies of a locally AdS$_3$ spacetime, namely $\mathcal{M}_1$ and $\mathcal{M}_2$ with coordinates $(t_1,z_1,x_1)$ and $(t_2,z_2,x_2)$, respectively. Each of these manifolds is further subdivided into two regions by the hypersurfaces $\Sigma_1$ and $\Sigma_2$ which are respectively given by $x_1=f_1(t_1,z_1)$ and $x_2=f_2(t_2,z_2)$. $\mathcal{M}_{1,L(R)}$ corresponds to $x_1 \leq(\geq)f_1(t_1,z_1)$ and similarly for $\mathcal{M}_2$. We glue $\mathcal{M}_{1,L}$ with $\mathcal{M}_{2,R}$ by identifying points on $\Sigma_1$ and $\Sigma_2$, such as (blue dots) $P_1$ and $P_2$ in the figure, which have the same worldsheet coordinates $\tau$ and $\sigma$. The gluing should satisfy the junction conditions with the stress tensor of a tensile string at the junction hypersurface resulting from the identification of $\Sigma_1$ and $\Sigma_2$.}
    \label{Fig:1}
\end{figure}

Consider two copies of a $2+1$-dimensional Einstein manifold $\mathcal{M}$, which is locally flat or an AdS space. We label the copies of $\mathcal{M}$  as $\mathcal{M}_1$ and $\mathcal{M}_2$. Let $\Sigma_1$ and $\Sigma_2$ be two $1+1$-dimensional hypersurfaces embedded in $\mathcal{M}_1$ and $\mathcal{M}_2$, respectively, and splitting each of these spacetimes into two halves. We will study solutions of Einstein's equations obtained by gluing one of the halves of $\mathcal{M}_1$ with one of the halves of $\mathcal{M}_2$ by identifying points in $\Sigma_1$ and $\Sigma_2$ such that the induced metrics and extrinsic curvatures satisfy the Israel junction conditions \cite{Israel:1966rt}. As mentioned before, our methodology is similar to that adopted in \cite{Bachas:2020yxv}, except that allow more general boundary conditions and also not restrict ourselves to linear perturbations about a simple solution. See Fig. \ref{Fig:1} for an illustration of the setup in AdS.

Let $t$, $z$ and $x$ be the coordinates of $\mathcal{M}$, where $t$ is the time coordinate, and $z$ and $x$ the spatial coordinates. The identical copies $\mathcal{M}_1$ and $\mathcal{M}_2$ of $\mathcal{M}$ are endowed with identical copies of the coordinate charts of $\mathcal{M}$, and the respective coordinates are $(t_1, z_1, x_1)$ and $(t_2, z_2, x_2)$. The hypersurface $\Sigma_1$ in $\mathcal{M}_1$ is given by
\begin{equation}
  x_1 = f_1(t_1, z_1),
\end{equation}
and we will call the half of $\mathcal{M}_1$ with $x_1 \leq f_1(t_1, z_1)$ as $\mathcal{M}_{1,L}$, and the other half as $\mathcal{M}_{1,R}$. Similarly, the hypersurface $\Sigma_2$ is given by
\begin{equation}
  x_2 = f_2(t_2, z_2),
\end{equation}
which splits $\mathcal{M}_2$ into $\mathcal{M}_{2,L}$ with $x_2 \leq f_2(t_2, z_2)$ and $\mathcal{M}_{2,R}$. We can glue either of the two halves of $\mathcal{M}_1$ with one of the halves of $\mathcal{M}_2$.  

In order to describe the gluing, it is convenient to fix a coordinate system $(\tau, \sigma)$ on the identified hypersurfaces $\Sigma_1$ and $\Sigma_2$ such that there is a pair of points, $P_1$ and $P_2$ belonging to $\Sigma_1$ and $\Sigma_2$, respectively, corresponding to each $(\tau, \sigma)$ (see Fig. \ref{Fig:1}). We use the following choice of the junction (worldsheet) coordinates -- for each $P_1 \equiv P_2$, we assign $(\tau, \sigma)$ such that 
\begin{equation}\label{Eq:WSGauge}
   \tau = \frac{1}{2}(t_1(P_1) + t_2(P_2)) , \quad \sigma = \frac{1}{2}(z_1(P_1) + z_2(P_2)).
\end{equation}
Having chosen this gauge fixing for the worldsheet diffeomorphisms, 
$\Sigma_1$  and $\Sigma_2$ take the parametric form 
\begin{align}
    &t_1(\tau, \sigma) = \tau - \tau_a(\tau, \sigma), \quad z_1(\tau, \sigma) = \sigma - \sigma_a(\tau, \sigma), \quad x_1(\tau, \sigma) = f_1\left(t_1(\tau, \sigma), z_1(\tau, \sigma)\right),\nonumber\\
    &t_2(\tau, \sigma) = \tau + \tau_a(\tau, \sigma), \quad z_2(\tau, \sigma) = \sigma + \sigma_a(\tau, \sigma), \quad x_2(\tau, \sigma) = f_2\left(t_2(\tau, \sigma), z_2(\tau, \sigma)\right).\label{Eq:Sigma12}
\end{align}
 Note $\tau_a = (1/2)(t_2 - t_1)$ and $\sigma_a =(1/2)(z_2 - z_1)$ are determined by the junction conditions, and not by the worldsheet gauge choice \eqref{Eq:WSGauge}. Together with $f_1$ and $f_2$, we thus obtain four functions which should be solved to obtain a gluing of one of the halves of $\mathcal{M}_{1}$ with one of the halves of $\mathcal{M}_{2}$. For later purposes, let us define
\begin{align}\label{Eq:xsxa}
    & x_s(\tau,\sigma) = \frac{1}{2}\left(f_1\left(t_1(\tau, \sigma), z_1(\tau, \sigma)\right)+f_2\left(t_2(\tau, \sigma), z_2(\tau, \sigma)\right)\right),\nonumber\\ & x_a(\tau,\sigma) = \frac{1}{2}\left(f_2\left(t_2(\tau, \sigma), z_2(\tau, \sigma)\right)-f_1\left(t_1(\tau, \sigma), z_1(\tau, \sigma)\right)\right),
\end{align}
so that $x_s$ is the average and $x_a$ is the relative $x$-coordinate of $\Sigma_1$ and $\Sigma_2$. 

For the sake of convenience, let us consider gluing $\mathcal{M}_{1,L}$ with $\mathcal{M}_{2,R}$ first.  In this case, the normals to both $\Sigma_1$ and $\Sigma_2$ are oriented along the directions in which the respective $x$-coordinate increases. Let $\gamma_{\mu\nu}^1$ and $\gamma_{\mu\nu}^2$ be the induced metrics, and $K_{\mu\nu}^1$ and $K_{\mu\nu}^2$ be the extrinsic curvatures of $\Sigma_1$ and $\Sigma_2$, respectively ($\mu$ and $\nu$ are worldsheet indices). Then the first junction condition
\begin{align}\label{Eq:gluing1}
    & \gamma_{\mu\nu}^1(\tau,\sigma) = \gamma_{\mu\nu}^2(\tau,\sigma) := \gamma_{\mu\nu}(\tau,\sigma), 
\end{align}
implies the continuity of the induced metric. The second junction condition states that the discontinuity of the Brown-York tensor $K_{\mu\nu}- K \gamma_{\mu\nu}$ should equal $8\pi G$ times the (conserved) stress tensor of the junction, which is $ T_0\gamma_{\mu\nu}$ if the junction consists of a string with a tension $T_0$ ($G$ is the three-dimensional Newton's constant). Let $\lambda = 8\pi G T_0$. It is easy to see that the trace-reversed second junction condition takes the following form,
\begin{align}\label{Eq:gluing2}
    & K_{\mu\nu}^1(\tau,\sigma) - K_{\mu\nu}^2(\tau,\sigma) = - \lambda \gamma_{\mu\nu}(\tau,\sigma).
\end{align}
Henceforth, we will refer to $\lambda$ as the tension. The two junction conditions give six equations. However, the Brown-York stress tensor is identically conserved on any hypersurface of an Einstein manifold by virtue of the Gauss-Codazzi equations and this guarantees that the traceless part of the Brown-York tensor is continuous at the junction even when $\lambda\neq 0$. Therefore, the gluing conditions \eqref{Eq:gluing2} give only one independent equation which together with \eqref{Eq:gluing1} amount to four equations for determining the four variables, namely $\tau_a$, $\sigma_a$, $f_1$ and $f_2$. 

 We will see that a solution of the junction conditions depends on the tension $\lambda$ and a finite number of integration constants $\delta_i$, that we call rigid deformations. The latter are associated with worldsheet isometries and rigid displacements of the hypersurface {via isometries of the embedding spacetime} which do not change its extrinsic curvature.

The main results we obtain in this paper are the following:

\begin{itemize}
    \item Given a solution of the junction conditions, 
    the hypersurface $\Sigma_{NG}$ in $\mathcal{M}$ given by
    \begin{equation}
    t= \tau, \quad z = \sigma, \quad x = x_s(\tau, \sigma; \lambda, \delta_i),
    \end{equation}
    satisfies an equation which reduces to the Nambu-Goto equation in the background $\mathcal{M}$ in the limit $\lambda, \delta_i \to 0$. We readily note that $\Sigma_{NG}$ has the average of the embedding coordinates of $\Sigma_1$ and $\Sigma_2$. Both the Nambu-Goto equation and the junction conditions (which give the deformed Nambu-Goto equation) can be solved with the same boundary and/or initial conditions for $x_s$, and this leads to a one-to-one correspondence between solutions $x_s(\lambda =0, \delta_i = 0)$ of the Nambu-Goto equation and solutions $x_s(\lambda, \delta_i)$ of the junction conditions. Generically, $x_s(\lambda,\delta_i) $ are smooth deformations of the corresponding Nambu-Goto solutions $x_s(\lambda =0, \delta_i = 0)$.  
    
    \item The other three variables $\tau_a$, $\sigma_a$ and $x_a$ are always determined uniquely by the Nambu-Goto solution $x_s(\lambda=0, \delta_i=0)$ for fixed values of $\lambda$ and $\delta_i$.
\end{itemize}
We will demonstrate the above results both in flat space and in AdS$_3$ using a perturbation expansion in which the tension is treated as a small parameter and we expand around solutions with zero tension. At the leading order, we choose an embedding for the junction which is a hypersurface with vanishing extrinsic curvature and which is also a simple solution of the Nambu-Goto equation.
The generic solutions of the Nambu-Goto equation appear at higher orders in the expansion. As a special case, when the  solution 
is the trivial one, we recover the non-linear static solution of \cite{Bachas:2021fqo} for $x_a$ which is determined solely by the tension.

If we glue $\mathcal{M}_{1,L}$ and $\mathcal{M}_{2,L}$ instead of gluing $\mathcal{M}_{1,L}$ and $\mathcal{M}_{2,R}$, we can use the same parametrizations in \eqref{Eq:Sigma12} for $\Sigma_1$ and $\Sigma_2$, but because of the change of orientation of the normal of $\Sigma_2$, the second gluing condition \eqref{Eq:gluing2} will be modified to 
\begin{align}\label{Eq:gluing4}
    & K_{\mu\nu}^1(\sigma, \tau) + K_{\mu\nu}^2(\sigma, \tau) = - \lambda \gamma_{\mu\nu}(\sigma, \tau).
\end{align}
The new solutions for the junction conditions are simply the same solutions for gluing $\mathcal{M}_{1,L}$ and $\mathcal{M}_{2,R}$ but with $x_s$ and $x_a$ interchanged. Therefore, in that case, $x_a$ instead of $x_s$ coincide with the solutions of Nambu-Goto equations when the rigid deformation parameters including the string tension vanish. One can similarly discuss the cases of gluing $\mathcal{M}_{1,R}$ and $\mathcal{M}_{2,R}$, and gluing $\mathcal{M}_{1,R}$ and $\mathcal{M}_{2,L}$.

Although the junction's stress tensor vanishes in the tensionless limit, it is not clear whether the spacetime is smooth in this limit. We need to construct coordinates in the bulk such that the tangents and the normal to the junction hypersurface are continuous to clarify this issue. This can be done in practice by utilizing the freedom of changing bulk coordinates on one side of the junction, but we leave this to the future. The discontinuities, if present, are determined by the rigid deformation parameters $\delta_i$ and the solution of the Nambu-Goto equation which fully characterize the junction.

\section{Junctions in flat space and the Nambu-Goto equation}\label{Sec:FS}
\subsection{Solving the junction conditions perturbatively}
We choose $\mathcal{M}$ to be Minkowski space $R^{2,1}$ endowed with the metric
\begin{equation}
    {\rm d}s^2 = - {\rm d}t^2 + {\rm d}z^2 + {\rm d}x^2.
\end{equation}
As described in the previous section, we consider two copies of $\mathcal{M}$ denoted by $\mathcal{M}_1$ and $\mathcal{M}_2$, each of which is split into two halves by the hypersurfaces $\Sigma_1$ and $\Sigma_2$, respectively. The parametric forms of these hypersurfaces are given in \eqref{Eq:Sigma12}. We then glue $\mathcal{M}_{1,L}$ and $\mathcal{M}_{2,R}$ with the junction conditions \eqref{Eq:gluing1} and \eqref{Eq:gluing2}. Here, we proceed with considering $\lambda$ to be small, i.e. $\lambda = \mathcal{O}(\varepsilon)$, so that we can determine the four functions $\tau_a$, $\sigma_a$, $f_1$ and $f_2$ by solving the junction conditions perturbatively in $\varepsilon$.

To set up the perturbative expansion, we choose $\Sigma_1$ and $\Sigma_2$ identically at the zeroth order such that their extrinsic curvatures vanish, while the continuity of the induced metric holds trivially at this order. This is accomplished by the choices 
\begin{equation}
    f_1  = x_0+ \mathcal{O}(\varepsilon),\quad f_2  = x_0+ \mathcal{O}(\varepsilon), \quad \tau_a = \sigma_a = \mathcal{O}(\varepsilon),
\end{equation}
so that $\Sigma_1$ and $\Sigma_2$ are the planes $x_1 = x_0$ and $x_2 = x_0$, respectively, at the zeroth order. To proceed systematically, the perturbative expansions of $\tau_a$ and $\sigma_a$ are taken to be  
\begin{align}\label{Eq:tausigmaexp}
    \tau_a(\tau, \sigma) = \sum_{k =1}^\infty \varepsilon^k \tau_{a,k}(\tau, \sigma), \quad \sigma_a(\tau, \sigma) = \sum_{k =1}^\infty \varepsilon^k \sigma_{a,k}(\tau, \sigma).
\end{align}
and similarly,
\begin{align}\label{Eq:f1f2exp}
    f_1 = x_0 + \sum_{k =1}^\infty \varepsilon^k f_{1,k}(\tau, \sigma), \quad f_2 = x_0+ \sum_{k =1}^\infty \varepsilon^k f_{2,k}(\tau, \sigma).
\end{align}
It is convenient to use the average and relative coordinates $x_s$ and $x_a$ as defined in \eqref{Eq:xsxa}. Obviously, the coefficients of their perturbative expansions are
\begin{align}\label{Eq:xsxaexp}
    x_{s,k}(\tau, \sigma) = \frac{1}{2} (f_{1,k}(\tau, \sigma)+f_{2,k}(\tau, \sigma)) , \quad  x_{a,k}(\tau, \sigma) = \frac{1}{2} (f_{2,k}(\tau, \sigma)-f_{1,k}(\tau, \sigma))
\end{align}
for $k=1,2,\cdots$, and at the zeroth order $x_{s,0} = x_0$ and $x_{a,0} =0$.

It turns out that $x_s$ is the only  propagating \textit{degree of freedom}. Therefore, we need to specify initial/boundary conditions for $x_s$ just like for a scalar field satisfying Klein-Gordon equations in $1+1$-dimensions. Here we will use an arbitrary length scale implicitly to make $\tau$ and $\sigma$ dimensionless. For the sake of illustration, we will choose these initial conditions for $x_s$ at $\tau=0$: 
\begin{equation}\label{Eq:IC0}
    x_{s}(0,\sigma) = x_0 +  A \sin \sigma, \quad \dot{x}_{s}(0,\sigma) = 0,
\end{equation}
where $\dot{f}$ denotes partial derivative of $f$ w.r.t. $\tau$. We will also assume that $A = \mathcal{O}(\varepsilon)$ so that the initial conditions for $x_{s,1}$ is 
\begin{equation}\label{Eq:IC1}
    x_{s,1}(0,\sigma) = A \sin \sigma, \quad \dot{x}_{s,1}(0,\sigma) = 0
\end{equation}
while the initial conditions for $x_{s,k}$ are 
\begin{equation}\label{Eq:IC2}
    x_{s,k}(0,\sigma)=\dot{x}_{s,k}(0,\sigma) =0, \quad {\rm for}\,\, k = 2,3,\cdots.
\end{equation}
We will find the most general solutions of the junction conditions corresponding to these initial conditions. These general solutions will have additional six rigid constant parameters as explained below.

As mentioned, the junction conditions are trivially satisfied at the zeroth order because
\begin{equation}
     \gamma^1_{\mu\nu} = \gamma^2_{\mu\nu}= \eta_{\mu\nu}+\mathcal{O}(\varepsilon), \quad K^1_{\mu\nu} = K^2_{\mu\nu} = \mathcal{O}(\varepsilon),
\end{equation}
where $\eta_{\mu\nu}= {\rm diag}(-1,1)$.

At the first order, the junction conditions \eqref{Eq:gluing1} for the continuity of the induced metric give
\begin{equation}\label{Eq:Junc11}
    \dot{\tau}_{a,1} = 0, \quad \dot{\sigma}_{a,1} - \tau^\prime_{a,1} = 0, \quad \sigma^\prime_{a,1} =0,
\end{equation}
 where $\dot{\phantom{a}}$ and $\phantom{a}'$ imply partial derivatives w.r.t. $\tau$ and $\sigma$, respectively. At this order, the junction conditions \eqref{Eq:gluing2} give
 \begin{equation}\label{Eq:Junc21}
    \ddot{x}_{a,1} = - \frac{\lambda}{2}, \quad \dot{x}^\prime_{a,1} = 0, \quad x^{\prime\prime}_{a,1} = \frac{\lambda}{2}.
\end{equation}
Notice that at this order, the junction conditions do not give an equation for $x_s$.
The solutions for $\tau_{a,1}$ and $\sigma_{a,1}$ are
\begin{equation}
    \tau_{a,1} = \betah + \alphah  \sigma, \quad \sigma_{a,1} = \gammah + \alphah \tau,
\end{equation}
which imply that the vector $\xi^\mu = (\tau_{a,1},\sigma_{a,1} )$ is a generic Killing vector on the worldsheet in the background of the zeroth order metric $\eta_{\mu\nu}$. We note that $\gammah$ and $\betah$ correspond to spacetime translations on the worldsheet while $\alphah$ is a worldsheet boost. The generic solution for $x_{a,1}$ is
\begin{equation}
    x_{a,1} = -\frac{\lambda}{4}(\tau^2 - \sigma^2) + \betak  + \alphak \sigma+ \gammak \tau.
\end{equation}
It is easy to see that $\betak$, $\alphak$ and $\gammak$ are rigid displacement parameters that preserve the extrinsic curvature of a hypersurface. {Explicitly, these correspond to infinitesimal isometries of the embedding flat space which do not keep the hypersurface $x=0$ (where $\Sigma_1$ and $\Sigma_1$ coincide at the zeroth order) invariant; $\alpha_k$ is a transverse spatial displacement, $\beta_k$ is an infinitesimal rotation and $\gamma_k$ is an infinitesimal transverse boost. The other three isometries, namely time translation, longitudinal spatial displacement and the longitudinal boost keep $x=0$ invariant.}

It will turn out that the higher orders in the expansion will not introduce new parameters, so that in total $\gammah$, $\betah$, $\alphah$, $\betak$, $\alphak$ and $\gammak$ give six rigid parameters which together with the initial conditions for $x_s$ uniquely specify a solution of the junction conditions to all orders. By assumption, all these six rigid parameters are small, i.e. $\mathcal{O}(\varepsilon)$. {Although these infinitesimal rigid parameters are related to worldsheet and spacetime isometries, at higher orders they generically affect the full non-linear solution non-trivially. Furthermore, these parameters should be determined by appropriate boundary conditions depending on the physical context, and therefore can generally be physical parameters. We will discuss boundary conditions explicitly in the context of junctions of AdS in the next section.}

\paragraph{Continuity of the induced metric at second and higher orders:} At second and higher orders, the junction conditions \eqref{Eq:gluing1} for the continuity of the induced metric take the following form
\begin{align}
    &\dot{\tau}_{a,n} - \left(\frac{\lambda}{2}\tau -\gammak\right) \dot{x}_{s,n-1}=\mathcal{A}^F_{n1}(\tau,\sigma),\label{Eq:junc1n1} 
    \\ &\dot{\sigma}_{a,n} - \tau^\prime_{a,n} +\left(\frac{\lambda}{2}\tau -\gammak\right) x^\prime_{s,n-1} -\left(\frac{\lambda}{2}\sigma +\alphak\right)\dot{x}_{s,n-1}= \mathcal{A}^F_{n2}(\tau,\sigma), \label{Eq:junc1n2} 
    \\ &\sigma^\prime_{a,n} -\left(\frac{\lambda}{2}\sigma +\alphak\right) x^\prime_{s,n-1} =\mathcal{A}^F_{n3}(\tau,\sigma),\label{Eq:junc1n3}
\end{align}
where $ \mathcal{A}^F_{n1}$, $ \mathcal{A}^F_{n2}$ and $ \mathcal{A}^F_{n3}$ are sources which are constituted of lower order terms, and also depend on the tension and the rigid parameters. These sources vanish for $n=2$. However, they are non-trivial for $n\geq 3$.

We use the following algorithm determine $\tau_{a,n}$, $\sigma_{a,n}$ and $x_{s,n-1}$ \textit{uniquely}:
\begin{itemize}
    \item We solve \eqref{Eq:junc1n1} and \eqref{Eq:junc1n3} first to obtain
    \begin{eqnarray}\label{Eq:tnxn}
    \sigma_{a,n}(\tau,\sigma) &=& h_{n1}(\tau)+ \alphak x_{s,n-1}(\tau,\sigma) \nonumber\\ &&\qquad+\frac{\lambda}{2}\int_0^\sigma {\rm d}\sigma_1 \,(\sigma_1  x^\prime_{s,n-1}(\tau, \sigma_1)+  \mathcal{A}^F_{n3}(\tau,\sigma_1)) ,\nonumber\\
    \tau_{a,n}(\tau,\sigma) &=& h_{n2}(\sigma)- \gammak x_{s,n-1}(\tau,\sigma)  \nonumber\\ &&\qquad+\frac{\lambda}{2}\int_0^\tau {\rm d}\tau_1\, (\tau_1  \dot{x}_{s,n-1}(\tau_1, \sigma)+ \mathcal{A}^F_{n1}(\tau_1,\sigma)).
    \end{eqnarray}
    \item Substituting \eqref{Eq:tnxn} in \eqref{Eq:junc1n2} we obtain
    \begin{align}\label{Eq:junc1n2p}
        & \dot{h}_{n1}(\tau) - h^\prime_{n2}(\sigma) + \frac{\lambda}{2}\left(\int_0^\sigma {\rm d}\sigma_1\,\sigma_1 \dot{x}^\prime_{s,n-1}(\tau, \sigma_1) - \int_0^\tau {\rm d}\tau_1\,\tau_1\dot{x}^\prime_{s,n-1}(\tau_1, \sigma)\right) \nonumber\\
    &-\frac{\lambda}{2}\left(\sigma \dot{x}_{s,n-1}(\tau, \sigma)-\tau x^\prime_{s,n-1}(\tau, \sigma)\right) =  \mathcal{\tilde S}^F_{n}(\tau,\sigma).
    \end{align}
    where $\mathcal{\tilde S}^F_{n}$ gets contribution from $ \mathcal{A}^F_{n1}$, $ \mathcal{A}^F_{n2}$ and $ \mathcal{A}^F_{n3}$.
    \item Differentiating \eqref{Eq:junc1n2p} first w.r.t. $\tau$ and then w.r.t $\sigma$ (or the other way around) we obtain
    \begin{equation}\label{Eq:xsn}
        \ddot{x}_{s,n-1}- {x}^{\prime\prime}_{s,n-1} = -\partial_\tau\partial_\sigma\mathcal{\tilde S}^F_{n}(\tau,\sigma).
    \end{equation}
    With the initial conditions \eqref{Eq:IC1} and \eqref{Eq:IC2} we obtain \textit{unique} solutions for $x_{s,n-1}$. As mentioned the source vanishes for $n=2$ yielding just the massless Klein-Gordon equation for $x_{s,2}$ on the worldsheet. It turns out that at the third order, $\partial_\tau\partial_\sigma\mathcal{\tilde S}^F_{3} =0$ (although $\mathcal{\tilde S}^F_{3}\neq 0$) implying that $x_{s,2}$ also follows the just the massless Klein-Gordon equation without any source term. We will show in section \ref{NGflat} that the equations \eqref{Eq:xsn} are just the deformations of perturbative expansion of the non-linear Nambu-Goto equations.
    \item Substituting the solution for $x_{s,n-1}$ in \eqref{Eq:junc1n2p} we get
    \begin{equation}\label{Eq:hn}
        \dot{h}_{n1}(\tau) - h^\prime_{n2}(\sigma) =0.
    \end{equation}
    whose solutions are 
    \begin{equation}
    h_{n1}(\tau) = \alpha_{hn} + \gamma_{hn}\tau, \quad h_{n2}(\sigma) = \beta_{hn} + \gamma_{hn}\sigma,
     \end{equation}
    where $\alpha_{hn}$, $\beta_{hn}$ and $\gamma_{hn}$ are constants. From \eqref{Eq:tnxn} it should be clear that these constants just modify the isometries given by $\tau_{a,1}$ and $\sigma_{a,1}$, and can be absorbed into $\alpha_{h}$, $\beta_{h}$ and $\gamma_h$, respectively. Therefore, these constants can be set to zero.
\end{itemize}

At the second order, the initial conditions \eqref{Eq:IC1} give 
\begin{equation}\label{Eq:xs1}
    x_{s,1}(\tau,\sigma) = A \sin\sigma \cos\tau
\end{equation}
and then we obtain explicitly that 
\begin{align}\label{Eq:x2t2}
    &\tau_{a,2}(\tau,\sigma)  = - A \gammak \cos\tau\sin\sigma +\frac{ A\lambda}{2} \left(\tau \cos\tau -\sin\tau\right)\sin\sigma,\nonumber\\
    &\sigma_{a,2}(\tau,\sigma)  =A \alphak \cos\tau\sin\sigma+\frac{ A\lambda}{2}\left(\sigma \sin\sigma +\cos\sigma\right)\cos\tau.
\end{align}
At the third and fourth orders, the initial conditions \eqref{Eq:IC2} give 
\begin{align}\label{Eq:xs3}
    & x_{s,2}(\tau,\sigma) = 0,\nonumber\\
    & x_{s,3}(\tau,\sigma) =\frac{1}{4} \Bigg(\frac{1}{2} A \sin \sigma  \Big(\tau  (2 A^2+(2
   \text{$\alphak $}+\lambda  \sigma )^2  \sin \tau  \nonumber\\
   &\quad +(\lambda  \tau -2 \text{$\gammak$})^2+2
   \lambda ^2)- (\frac{A^2}{4}+2 \lambda ^2 \tau
   ^2)\cos \tau+\frac{1}{4} A^2 \cos (3 \tau )\Big)
   \nonumber\\&+\frac{1}{2} A^3 \sin (3 \sigma )
   \cos \tau(\sin \tau)^2   \nonumber\\&
   -A (2 \text{$\alphak$}+\lambda  \sigma )\cos \sigma  
   \left( (2 \text{$\gammak$}+\lambda  \tau )\sin \tau +\tau(\lambda 
   \tau -2 \text{$\gammak$})  \cos \tau \right) \Bigg),
\end{align}
and so on.

 It is important to point out that we cannot obtain the Nambu-Goto equations if we directly work with $\lambda = 0$ instead of taking the limit $\lambda\rightarrow 0$. We do not obtain the equations for $x_{s,n}$ given by \eqref{Eq:xsn} from \eqref{Eq:junc1n2p} when $\lambda = 0$. In this case, the sources $\mathcal{\tilde S}^F_{n}$ vanish and we obtain \eqref{Eq:hn} directly from \eqref{Eq:junc1n2p} so that $\tau_{a,n}$ and $\sigma_{a,n}$ are given by $x_{s,n-1}$ and lower order terms, and $x_s$ as expected can be chosen arbitrarily.\footnote{At least when $x_a = 0$, $\Sigma_1$ and $\Sigma_2$ can coincide on any arbitrary hypersurface giving back the full spacetime after gluing when we set $\lambda = 0$.} On the other hand, if we take the limit $\lambda \rightarrow 0$, $x_{s,n}$ coincides with a rigid deformation of a perturbative solution of the Nambu-Goto equation.

\paragraph{Discontinuity of the extrinsic curvature at second and higher orders:} The other set of junction conditions \eqref{Eq:gluing2} relating the discontinuity of the extrinsic curvature to the stress tensor of the junction yield the equations which determine $x_{a,n}$ for $n\geq 2$. Schematically, these equations are
\begin{equation}\label{Eq:Junc22}
    \ddot{x}_{a,n} = \mathcal{B}^F_{n1}(\tau,\sigma), \quad \dot{x}^{\prime}_{a,n} = \mathcal{B}^F_{n2}(\tau,\sigma), \quad {x}^{\prime\prime}_{a,n} = \mathcal{B}^F_{n3}(\tau,\sigma),
\end{equation}
where $\mathcal{B}^F_{n1}$, $\mathcal{B}^F_{n2}$ and $\mathcal{B}^F_{n3}$ are sources determined by lower order terms and $\lambda$. These equations are of the same form as \eqref{Eq:Junc21} which appear in the first order except for more complicated source terms. As already discussed, the condition \eqref{Eq:gluing2} gives only one independent equation. This implies that the source terms are not independent. Our general strategy for solving these equations is as follows:
\begin{itemize}
    \item Use the last equation in \eqref{Eq:Junc22} above to obtain
    \begin{equation}
        x_{a,n} = k_{n1}(\tau) + k_{n2}(\tau)\sigma + \int_0^\sigma {\rm d}\sigma_1 \int_0^{\sigma_1} {\rm d}\sigma_2\, \mathcal{B}^F_{n3}(\sigma_2, \tau)
    \end{equation}
    \item Substituting the above form of $x_{a,n}$ into the second equation in \eqref{Eq:Junc22} simply yields
    \begin{equation}
        \dot{k}_{n2}(\tau) = F_{n1}(\tau),
    \end{equation}
    with $F_{n1}(\tau)$ derived from $\mathcal{B}^F_{n2}(\tau,\sigma)$ and $\mathcal{B}^F_{n3}(\tau,\sigma)$, and thus 
    \begin{equation}
        k_{n2}(\tau) = \beta_{kn} + \int_0^\tau {\rm d}\tau_1 F_{n1}(\tau_1).
    \end{equation}
    This statement is far from obvious since both $\mathcal{B}^F_{n2}$ and $x_{a,n}$ depend on both $\tau$ and $\sigma$. However, as noted before, due to the conservation of the Brown-York tensor on any hypersurface in an Einstein manifold, the junction conditions \eqref{Eq:gluing2} are not independent of each other.
    \item Finally substituting the above form of $x_{a,n}$ and $k_{n2}$ into the first equation in \eqref{Eq:Junc22} simply yields
    \begin{equation}
        \ddot{k}_{n1}(\tau) = F_{n2}(\tau),
    \end{equation}
    with $F_{n2}(\tau)$ derived from $\mathcal{B}^F_{n2}(\tau,\sigma)$ and $\mathcal{B}^F_{n3}(\tau,\sigma)$, and thus 
    \begin{equation}
        k_{n2}(\tau) = \alpha_{kn} + \gamma_{kn}\tau   + \int_0^\tau {\rm d}\tau_1 \int_0^{\tau_1} {\rm d}\tau_2\, F_{n2}(\tau_2).
    \end{equation}
    Once again this statement is far from obvious but fundamentally is also a result of the consistency of the junction conditions.
    \item 
    Finally, we set $\alpha_{kn}$, $\beta_{kn}$ and $\gamma_{kn}$ to zero as these can be absorbed into $\alpha_k$, $\beta_{k}$ and $\gamma_{k}$, respectively which appear at the first order and give generic rigid deformations of a hypersurface which do not change its extrinsic curvature. Thus we fully determine $x_{a,2}$.
\end{itemize}
At the second order, we obtain
\begin{equation}
    x_{a,2} = A (\alpha_h + \alphah \tau)\cos\tau\cos\sigma -  A (\beta_h + \alphah \sigma)\sin\tau\sin\sigma,
\end{equation}
and so on. We can immediately note from the form of $x_{a,2}$ that the tensionless limit $\lambda\rightarrow 0$ is non-trivial since $x_{a,2}$ is actually independent of the tension like $x_{a,1}$. 

We also readily note that the perturbative expansion breaks down at large $\vert \tau \vert$ and large $\vert \sigma \vert$. 
Therefore, we cannot determine the asymptotic behavior of the junction; this issue could perhaps be addressed by studying the problem in the Bondi coordinates of $\mathcal{M}$. We leave this to the future.

\subsection{Matching with the solutions of the Nambu-Goto equation}\label{NGflat}

Let us consider the hypersurface $\Sigma_{NG}$ in $\mathcal{M}$ whose parametric form is
\begin{equation}
    t = \tau, \quad z = \sigma, \quad x = f(\tau, \sigma).
\end{equation}
The Nambu-Goto equation for this hypersurface in three-dimensional Minkowski space is 
\begin{equation}\label{Eq:NGFS}
    \ddot{f}\left(1+ {f^\prime}^2\right)-f^{\prime\prime}\left(1- \dot{f}^2\right) - 2 \dot{f}f^\prime \dot{f}^\prime = 0.
\end{equation}
We can solve this perturbatively via the following expansion in $\varepsilon$ (the amplitude of the linearized perturbation):
\begin{equation}\label{Eq:fpert}
    f(\tau,\sigma) = x_0 + \sum_{k=1}^\infty \varepsilon^k f_k(\tau, \sigma).
\end{equation}
In order to examine the correspondence with the gravitational spacetime with junction, we use the same initial conditions \eqref{Eq:IC0} for $x_s$, i.e. 
\begin{equation}\label{Eq:IC0f}
    f(0,\sigma) = x_0 + A \sin\sigma, \quad \dot{f}(0,\sigma) = 0.
\end{equation}
Assuming that $A$ is $\mathcal{O}(\varepsilon)$, the above amounts to
\begin{equation}\label{Eq:IC1f}
    f_{1}(0,\sigma) = A \sin \sigma, \quad \dot{f}_{1}(0, \sigma) = 0
\end{equation}
and
\begin{equation}\label{Eq:IC2f}
    f_{k}(0, \sigma)=\dot{f}_{k}(0,\sigma) =0, \quad {\rm for}\,\, k = 2,3,\cdots,
\end{equation}
which coincide with the initial conditions for $x_{s,k}$ as given in \eqref{Eq:IC1} and \eqref{Eq:IC2}. 

Since the Nambu-Goto equation \eqref{Eq:NGFS} is odd in $f$, clearly $f_{n}$ vanishes when $n$ is a positive even integer. Else we obtain,
\begin{equation}\label{Eq:fn}
    \ddot{f}_1 - f^{\prime\prime}_1 = 0, \quad \ddot{f}_{2n +1} - f^{\prime\prime}_{2n +1} = \mathcal{S}^F_{2n+ 1}, \,\, {\rm for}\,\, n\geq 1,
\end{equation}
where $\mathcal{S}^F_{2n+1}$ are sources constituted of lower order terms.

At the first order, the initial conditions \eqref{Eq:IC1f} give 
\begin{equation}
    f_{1} = A \sin\sigma\cos\tau.
\end{equation}
We note that $f_1$ is identical with $x_{s,1}$ which given by \eqref{Eq:xs1}. At the third order, the initial conditions \eqref{Eq:IC2f} give
\begin{align}
    f_{3} = \frac{1}{2}A^3\sin\sigma\sin\tau(2\tau + \cos(2\sigma)\sin(2\tau)),
\end{align}
and so on. Using trigonometric identities, we can check that $f_3$ coincides with $x_{s,3}$ (given by \eqref{Eq:xs3}) when $\lambda$, $\alpha_k$, $\beta_k$ and $\gamma_k$ vanish.

Generally, we can verify that when $\lambda$ and the six rigid parameters $\alpha_h$, $\beta_h$, $\gamma_h$, $\alpha_k$, $\beta_k$ and $\gamma_k$ vanish, then $x_{s,n}$ vanish for positive even integral values of $n$, while $x_{s,n}$ coincides with $f_{n}$ for positive odd integral values of $n$. This implies that $f$ coincides with $x_s$ when $\lambda$ and the rigid deformations vanish. Thus when we set identical initial conditions for $x_s$ and $f$, there is a one-to-one correspondence between the solutions of the junction conditions and the Nambu-Goto equations up to the six rigid deformations.

We emphasize that the specific choice of initial conditions is not important to show this correspondence since the perturbative expansions of the Nambu-Goto equation \eqref{Eq:fn} themselves coincide with the equations for $x_{s,n}$ given by \eqref{Eq:xsn} when the string tension and the rigid parameters vanish. As for illustration, the source terms appearing in \eqref{Eq:xsn} and \eqref{Eq:fn} at the first non-trivial order are 
\begin{align}
  -\partial_{\tau} \partial_{\sigma}\mathcal{\tilde S}^F_{4} :\equiv&  ~ 4 A^3 \cos{\tau}\sin{\sigma} \left(2- \cos{ 2\tau} + \cos{2\sigma}\right)\nonumber \\ &+ 2 A \left(\lambda \sigma-2 \gamma_k \right)\sin{\tau} \left(2 \beta_k \cos{\sigma} +\lambda (\sigma \cos{\sigma}+2\sin{\sigma})\right)\nonumber \\
  &+\frac{A}{2}\cos{\tau}\left(-4\lambda (2 \beta_k +\lambda \sigma) \cos{\sigma} + \left((\lambda \tau- 2 \gamma_k)^2 +(\lambda \sigma+ 2 \beta_k)^2\right)\sin{\sigma}\right),\nonumber\\
  \mathcal{S}^{F}_{3} : \equiv & ~4 A^3 \cos{\tau}\sin{\sigma} \left(2- \cos{ 2\tau} + \cos{2\sigma}\right). \nonumber
\end{align}
which source $x_{s,3}$ and $f_3$, respectively. Clearly these sources agree in the limit $\lambda \rightarrow 0, \beta_k \rightarrow 0 , \gamma_k \rightarrow 0$.

\section{Junctions in anti-de Sitter space and the Nambu-Goto equation}\label{Sec:AdS}

\subsection{Solving the junction conditions perturbatively}
Here we will study the case of gluing two identical copies of a locally AdS$_3$ spacetime $\mathcal{M}$ across a stringy junction. As in the case of flat space, we will glue $\mathcal{M}_{1.L}$ with $\mathcal{M}_{2.R}$.

Especially motivated by applications to the holographic correspondence and to the understanding of black holes, we consider the locally AdS$_3$ space to be a Ba{\~n}ados-Teitelboim-Zanelli (BTZ) black hole \cite{Banados:1992wn,Banados:1992gq} with a finite mass $M$. For simplicity, we consider the case of vanishing (angular) momentum. We proceed with the choice of units in which the radius of AdS$_3$ is unity and the following convenient coordinates in which the metric of the BTZ black hole is of the form:
\begin{equation}\label{Eq:BTZm}
    {\rm d}s^2 = \frac{{\rm d}z^2}{z^2 - M}- (z^2 - M) {\rm d}t^2 + z^2  {\rm d}x^2.
\end{equation}
Note that $z$ is the radial coordinate and the spacetime boundary is at $z= \infty$.  There is a coordinate singularity at the horizon $z=\sqrt{M}$. Note that $x$ is not periodic so the above is the metric for the black hole in Poincar\'{e} AdS$_3$.

We set up the perturbative expansion exactly like in the case of flat space with the tension $\lambda = \mathcal{O}(\varepsilon)$. We choose $\Sigma_1$ and $\Sigma_2$ to be the hypersurfaces $x_1 = x_0$ and $x_2 = x_0$, respectively at the zeroth order, and we choose the worldsheet gauge to be \eqref{Eq:WSGauge} so that the average time coordinate $t_{s} = 1/2(t_1 + t_2)$ and the average radial coordinate $z_s = 1/2(z_1 + z_2)$ coincide with $\tau$ and $\sigma$, respectively. At the zeroth order, the induced metrics on the hypersurfaces $\Sigma_1$ and $\Sigma_2$ are trivially identical. Explicitly,
\begin{align}\label{Eq:WSM0}
    &\gamma^1_{\tau\tau} = \gamma^2_{\tau\tau} = - (\sigma^2 -M) + \mathcal{O}(\varepsilon), \quad \gamma^1_{\tau\sigma} = \gamma^2_{\tau\sigma} =\mathcal{O}(\varepsilon),\nonumber\\ &\gamma^1_{\sigma\sigma} = \gamma^2_{\sigma\sigma} = \frac{1}{\sigma^2-M}+ \mathcal{O}(\varepsilon).
\end{align}
This induced metric at the leading order is locally AdS$_2$. It actually describes a AdS$_2$ black hole with horizon at $\sigma = \sqrt{M}$ and boundary at $\sigma = \infty$. Furthermore, the extrinsic curvatures $K^1_{\mu\nu}$ and $K^2_{\mu\nu}$ vanish at the leading order. The perturbative expansions for $\tau_a$ and $\sigma_a$ are given by \eqref{Eq:tausigmaexp}, while those of $f_1$ and $f_2$ are given by \eqref{Eq:f1f2exp}. Obviously, $x_{s,0} = x_0$, $x_{a,0}=0$, and at higher order expansions of $x_s$ and $x_a$ are given by \eqref{Eq:xsxaexp}.

Once again $x_s$ will be the only propagating degree of freedom emerging from solving the junction conditions. To specify the solutions for $x_s$, we need to specify boundary conditions just like for a scalar field in AdS$_2$. We impose Dirichlet boundary conditions at $\sigma=\infty$ for $x_s$ so that we can readily compare the solutions of $x_s$ with those of the Nambu-Goto equation for a string in $\mathcal{M}$ in which the endpoint of the string is pinned to a fixed value at the boundary of $\mathcal{M}$. Particularly, this boundary condition implies that
\begin{equation}\label{Eq:DBCxs}
    \lim_{\sigma \rightarrow\infty}x_{s} = x_0, \quad {\rm i.e.}, \quad\lim_{\sigma \rightarrow\infty} x_{s,i} = 0, \,\,{\rm for}\,\, i = 1,2, \cdots,
\end{equation}
since $x_{s,0} = x_0$ identically. Furthermore, we also demand that $x_s$ and therefore each $x_{s,i}$ individually satisfy the ingoing boundary condition at the worldsheet horizon $\sigma = \sqrt{M}$. These boundary conditions are also natural when we compare these solutions with the solutions of the Nambu-Goto equations. In both cases, we will obtain the same spectrum of quasi-normal modes and there will be a one-to-one correspondence between the two solutions even at the non-linear level up to rigid deformations. Although here we will restrict our analysis to the ingoing boundary condition at the horizon and Dirichlet boundary condition at the boundary of AdS, we expect the correspondence to be valid for other boundary conditions as well.

As in the case of flat space, we study general solutions of the junction conditions with the above boundary conditions for $x_s$. We find that these solutions have additional finite number of rigid parameters. However, in the  case of AdS, it is well motivated to impose the Dirichlet boundary conditions on both $x_1$ and $x_2$ so that the boundary spacetime remains unmutilated; therefore  we require that 
\begin{equation}\label{Eq:DBCx1x2}
    \lim_{\sigma \rightarrow\infty}x_1 = x_0, \quad {\rm i.e.}, \quad\lim_{\sigma \rightarrow\infty} x_2 = x_0.
\end{equation}
This implies that the endpoints of both $\Sigma_1$ and $\Sigma_2$ at the boundary of $\mathcal{M}_1$ and $\mathcal{M}_2$, respectively, are pinned to the same constant value $x_0$.  To implement this, together with the Dirichlet boundary conditions \eqref{Eq:DBCxs} on $x_s$ we need to impose the following Dirichlet boundary condition for $x_a$:
\begin{equation}\label{Eq:DBCxa}
    \lim_{\sigma \rightarrow\infty}x_a = 0, \quad {\rm i.e.}, \quad\lim_{\sigma \rightarrow\infty} x_{a,i} = 0,  \,\,{\rm for}\,\, i = 1,2, \cdots,
\end{equation}
since $x_{a,0} =0$ identically. It is not at all obvious that such boundary conditions can be implemented since $x_{a}$, unlike $x_s$ is not a degree of freedom. However, as discussed below, we will see that the boundary conditions \eqref{Eq:DBCxs} and \eqref{Eq:DBCxa} can be simultaneously imposed if we reduce the number of possible rigid deformations.

One crucial point is that we do not impose any boundary conditions on $\tau_{a}$ and $\sigma_a$. In fact, these will be given by two rigid parameters related to worldsheet isometries at the leading order.\footnote{This is in contrast to the setup of \cite{Bachas:2020yxv}, where they imposed Dirichlet boundary conditions on all the variables and found a unique solution for each given frequency in pure AdS.} The diffeomorphisms related to these two parameters preserve the worldsheet boundary at $\sigma =\infty$ but induce relative time-reparametrization at the junction since $\tau_{a}$ does not vanish at the boundary of AdS.

Unlike the case of flat space, our perturbative expansion will be valid near the boundary although it will be unreliable near the horizon.

At the zeroth order, the junction conditions are trivially satisfied just like in the case of flat space discussed previously. At the first order, the conditions \eqref{Eq:gluing1} for the continuity of the induced metric at the junction gives
\begin{align}\label{Eq:hcont1}
   \dot{\tau}_{a,1} +\frac{\sigma}{\sigma^2-M} \sigma_{a,1} =0,\nonumber\\
   \tau^{\prime}_{a,1} -\frac{1}{(\sigma^2 - M)^2} \dot{\sigma}_{a,1} =0,\nonumber\\
  \sigma^{\prime}_{a,n} -\frac{\sigma}{\sigma^2 - M} \sigma_{a,n} =0,
\end{align}
while the other set of junction conditions \eqref{Eq:gluing2} relating the discontinuity of the extrinsic curvature at the junction to the worldsheet stress tensor gives
\begin{align}\label{Eq:Kdisc1}
  \ddot{x}_{a,1} - \sigma(\sigma^2 - M)x^\prime_{a,1}=-\frac{\lambda (\sigma^2 - M)}{2\sigma},\nonumber\\
 \dot{x}^{\prime}_{a,1} -\frac{M\dot{x}_{a,1}}{\sigma(\sigma^2 - M)} =0,\nonumber\\
    x^{\prime\prime}_{a,1} +\frac{3\sigma^2 -2M}{ \sigma(\sigma^2-M)} x^\prime_{a,1} =\frac{\lambda}{2\sigma(\sigma^2-M)}.
\end{align}
The general solutions for \eqref{Eq:hcont1} are
\begin{align}\label{Eq:t1x1AdS}
    &\tau_{a,1} = \alpha_h + \frac{\sigma}{\sqrt{\sigma^2 - M}}\left(-\beta_h e^{\sqrt{M}\tau}+\gamma_h e^{-\sqrt{M}\tau}\right), \nonumber\\
    &\sigma_{a,1} = \sqrt{M(\sigma^2 - M)}\left(\beta_h e^{\sqrt{M}\tau}+\gamma_h e^{-\sqrt{M}\tau}\right),
\end{align}
and it is easy to check that $\xi^\mu = (\tau_{a,1},\sigma_{a,1})$ is a generic Killing vector in the zeroth order background metric, which in this case is the locally AdS$_2$ metric \eqref{Eq:WSM0}. The parameters $\alpha_h$, $\beta_h$, and $\gamma_h$  therefore correspond to the (infinitesimal) generators of the $SL(2,R)$ isometries. We note that although the boundary at $\sigma =\infty$ is preserved, the time is reparametrized at the boundary. The general solutions of \eqref{Eq:Kdisc1} are
\begin{equation}\label{Eq:xa1AdS}
    x_{1,a} = - \frac{\lambda}{2\sigma} + \alpha_{k} + \frac{\sqrt{\sigma^2 - M}}{\sigma}\left(\beta_k e^{\sqrt{M}\tau}+\gamma_k e^{-\sqrt{M}\tau}\right)
\end{equation}
Just like in the case of flat space, the parameters $\alpha_k$, $\beta_k$, and $\gamma_k$ parametrize the rigid infinitesimal deformations of a hypersurface which preserves its extrinsic curvature. {As in the context of flat space, these correspond to infinitesimal isometries of the background metric which do not preserve the zeroth order hypersurface $x=0$.}\footnote{Explicitly, the six Killing vectors of the metric \eqref{Eq:BTZm} with parameters $a_1$, $a_2$, $a_3$, $a_4$, $b$ and $c$ are 
\begin{equation*}
    \xi^z = \sqrt{z^2 - M}\, a(t,x), \quad \xi^t = b -\frac{z}{M\sqrt{z^2 - M}}\partial_t a(t,x), \quad \xi^x = c -\frac{\sqrt{z^2 - M}}{z M}\partial_x a(t,x),
\end{equation*}
with 
\begin{equation*}
    a(t,x) = a_1 e^{\sqrt{M}(x+t)} + a_2 e^{\sqrt{M}(x-t)}+a_3 e^{-\sqrt{M}(x+t)} + a_4 e^{-\sqrt{M}(x-t)}.
\end{equation*}
Comparing \eqref{Eq:xa1AdS} with $\xi^x$ after setting $z=\sigma$ (as at the zeroth order) and $x=0$, we find that $\alpha_k$, $\beta_k$ and $\sigma_k$ correspond to the isometries which do not preserve the hypersurface $x=0$.
} {The rigid parameters affect the generic full solution of the junction at higher orders non-trivially. Generally, the boundary conditions can determine some of the rigid parameters, and the remaining ones could physically characterize the nature of the interface dual to the bulk junction. We will comment more on this in the concluding section.}

Specifically, imposing the Dirichlet boundary condition on both $x_1$ and $x_2$, leads to the boundary conditions \eqref{Eq:DBCxa} for $x_{a,1}$ implying that it should vanish at $\sigma\rightarrow\infty$. This can be satisfied if 
\begin{equation}
  \alpha_k = \beta_k = \gamma_k =0,
\end{equation}
We will find that the junction conditions can be solved perturbatively only when the exponentially growing modes $\beta_h$ and $\beta_k$ are set to zero when we impose ingoing boundary conditions for $x_s$ at the worldsheet horizon. We will proceed below with imposing Dirichlet boundary condition on $x_s$ only to keep our discussion more general and so we will not set $\alpha_k$, $\beta_k$, and $\gamma_k$ to zero. Later we will impose the Dirichlet boundary condition on $x_a$ as well.

\paragraph{Continuity of the induced metric at second and higher orders:} At second and higher orders, the continuity of the induced metric gives
\begin{align}
  &  \dot{\tau}_{a,n} +\frac{\sigma}{\sigma^2-M} \sigma_{a,n}-\frac{\sqrt{M}\sigma}{\sqrt{\sigma^2 - M}}(\gamma_k e^{-\sqrt{M}\tau}- \beta_k e^{\sqrt{M}\tau}) \dot{x}_{s,n-1} =\mathcal{A}^A_{n1},\label{Eq:hcontn1}\\
 & \tau^{\prime}_{a,n} -\frac{1}{(\sigma^2 - M)^2}\dot{\sigma}_{a,n}+\frac{1}{\sigma^2 - M}\left(\frac{\lambda}{2}+\frac{e^{\sqrt{M}\tau}M}{\sqrt{\sigma^2 - M}}\beta_k+\frac{e^{-\sqrt{M}\tau}M}{\sqrt{\sigma^2 - M}}\gamma_k\right) \dot{x}_{s,n-1}\nonumber\\& + \frac{\sqrt{M}\sigma}{\sqrt{\sigma^2 - M}}\left(e^{\sqrt{M}\tau}\beta_k - e^{-\sqrt{M}\tau}\gamma_k \right){x}^\prime_{s,n-1}=\mathcal{A}^A_{n2},\label{Eq:hcontn2}\\
  & \sigma^{\prime}_{a,n} -\frac{\sigma}{\sigma^2 - M} \sigma_{a,n} \nonumber\\
   &-\frac{1}{2}(\sigma^2 - M)^{1/2}\left(\lambda \sqrt{\sigma^2 - M}+2 M e^{-\sqrt{M}\tau}\gamma_k+2 M e^{\sqrt{M}\tau}\beta_k\right)x^\prime_{s,n-1}=\mathcal{A}^A_{n3}\label{Eq:hcontn3}.
\end{align}
where the sources $\mathcal{A}^A_{ni}$ are constituted of lower order terms. The second order sources $\mathcal{A}^A_{2i}$ vanish. We determine $\tau_{a,n}$, $\sigma_{a,n}$, and $x_{s,n-1}$ from these equations following the algorithm mentioned below which is very similar to the case of flat space. It is based on the nested structure of these equations (it may be helpful for the reader to recall the discussion in the previous section).
\begin{itemize}
   \item We first solve \eqref{Eq:hcontn3} to obtain $\sigma_{a,n}$ and substitute this form of $\sigma_{a,n}$ into \eqref{Eq:hcontn1} to determine $\tau_{a,n}$. From these equations, we can readily see that we determine these up to two additive functions of only $\sigma$ and only $\tau$ respectively. Explicitly, we obtain
    \begin{align}\label{Eq:tnxnAdS}
        &\tau_{a,n} = s_{1n}(\tau, \sigma) - \frac{\sigma}{\sqrt{\sigma^2 -M}} h_{n1}(\tau) + h_{n2}(\sigma),\nonumber\\
        &\sigma_{a,n} = s_{2n}(\tau, \sigma) + \sqrt{\sigma^2 - M}\,\dot{h}_{n1}(\tau),
    \end{align}
    where $s_{1n}$ and $s_{2n}$ are determined fully by $x_{s,n-1}$ and lower order terms while $h_{n1}(\tau)$ and $h_{n2}(\sigma)$ are undetermined functions of $\tau$ and $\sigma$, respectively.
    \item When the above forms of $\sigma_{a,n}$ and $\tau_{a,n}$ are substituted in \eqref{Eq:hcontn2}, the latter assumes the form we call $\mathcal{E}$ (of course $\mathcal{E}$ should vanish). Similarly to the case of flat space, we find that $(\dot{\mathcal{E}}(\sigma^2 - M)^{3/2})^\prime = 0$ implies a differential equation for $x_{s,n-1}$ which takes the form:
    \begin{align}\label{Eq:xsAdSn}
       \frac{1}{\sigma^2 -M} \ddot{x}_{s,n-1} - (\sigma^2 -M){x}^{\prime\prime}_{s,n-1} -\frac{2(2\sigma^2 -M)}{\sigma}x^\prime_{s,n-1} =\widetilde{\mathcal{S}}^A_{n},
    \end{align}
    where $\widetilde{\mathcal{S}}^A_{n}$ is determined by $\mathcal{A}^A_{n2}$, $s_{1n}$ and $s_{2n}$. As discussed in the next subsection, the above matches \textit{exactly} with the perturbative expansion of the Nambu-Goto equation in $\mathcal{M}$ when the rigid parameters and $\lambda$ vanish. For $n=2$ and $n=3$, the source vanishes and the above results in identical homogeneous equations for $x_{s,1}$ and $x_{s,2}$ which match exactly with the linearized Nambu-Goto equation in $\mathcal{M}$.  
    
    \item Finally, when $x_{s,n-1}$ satisfies \eqref{Eq:xsAdSn}, we find that \eqref{Eq:hcontn2} (i.e. $\mathcal{E}$) reduces to 
    \begin{equation}\label{Eq:hneqAdS}
        \ddot{h}_{n1} - M h_{n1}(\tau) - (\sigma^2 - M)^{\frac{3}{2}}h^\prime_{n2}(\sigma) = K_n(\tau,\sigma).
    \end{equation}
    This equation is not guaranteed to have a solution. However, when the exponential growing modes $\beta_h$ and $\beta_k$ vanish, it turns out that 
    \begin{equation}
        K_n(\tau,\sigma) = F_n(\tau),
    \end{equation}
    a function of $\tau$ only. Then we obtain that
    \begin{align}\label{Eq:hnAdS}
        &h_{n1}(\tau) = \beta_{h,n} e^{\sqrt{M}\tau} + \gamma_{h,n} e^{-\sqrt{M}\tau} \nonumber\\&
        + \frac{e^{\sqrt{M}\tau} }{2\sqrt{M}}\int_0^\tau {\rm d}\tau_1 \, e^{-\sqrt{M}\tau_1} F_n(\tau_1) -\frac{e^{-\sqrt{M}\tau} }{2\sqrt{M}}\int_0^\tau {\rm d}\tau_1 \, e^{\sqrt{M}\tau_1} F_n(\tau_1),\nonumber\\
        & h_{n2}(\sigma) = \alpha_{h,n}.
    \end{align}
    We readily see by comparing \eqref{Eq:tnxnAdS} with \eqref{Eq:t1x1AdS} that the parameters $\alpha_{h,n}$, $\beta_{h,n}$ and $\gamma_{h,n}$ can be respectively absorbed into $\alpha_h$, $\beta_h$ and $\gamma_h$, the infinitesimal generators of the $SL(2,R)$ isometries of the zeroth order locally AdS$_2$ worldsheet metric \eqref{Eq:WSM0}. Therefore, we set $\alpha_{h,n}$, $\beta_{h,n}$ and $\gamma_{h,n}$ to zero for $n\geq 2$. Note that $\beta_h$ and $\beta_k$ should also vanish for \eqref{Eq:hneqAdS} to have solutions as mentioned above.
\end{itemize}
Thus the above algorithm determine the solutions for $\tau_{a,n}$ and $\sigma_{a,n}$ for $n\geq 2$, and $x_{s,n}$ for $n\geq 1$ up to four rigid deformation parameters, namely $\alpha_h$, $\gamma_h$, $\alpha_k$ and $\gamma_k$ since solutions of \eqref{Eq:hcontn1}, \eqref{Eq:hcontn2} and \eqref{Eq:hcontn3} exist only when $\beta_h =\beta_k = 0$. The latter is an artefact of choosing in-going boundary conditions for $x_{s,n}$ at the worldsheet horizon. If we would have chosen outgoing solutions instead, then we would have needed to set $\gamma_h =\gamma_k = 0$. Thus the perturbative expansion breaks down either in the far past or in the far future, and we choose $\beta_h =\beta_k = 0$ so that the expansion works in the far future.  

As mentioned above, $x_{s,n}$ satisfies the Dirichlet boundary conditions \eqref{Eq:DBCxs} at the boundary of the worldsheet and also the ingoing boundary condition at the worldsheet horizon. For $n=1$, \eqref{Eq:xsAdSn} is a homogeneous equation whose general solutions satisfying these boundary conditions are: 
\begin{equation}\label{solxs1}
    x_{s,1}(\tau,\sigma) = \sum_{n=0}^\infty A_n e^{-(2+n)\sqrt{M}\tau} \,\sigma^{-1} Q_1^{2+n}\left(\dfrac{\sigma}{\sqrt{M}}\right),
\end{equation}
where $Q_p^q(x)$ is an associated Legendre function of the second kind. Explicitly,
\begin{equation}
    Q_1^2(x) =  \frac{2}{1 -x^2},\quad Q_1^3(x) = - \frac{8x}{(1-x^2)^{3/2}}, \quad  Q_1^4(x) = \frac{8(1+5x^2)}{(1-x^2)^2}, \cdots.
\end{equation}
Note that $A_n$ are all $\mathcal{O}(\varepsilon)$ and are defined with phases so that $x_{s,1}$ is real. The general linear solution \eqref{solxs1} is a superposition of quasi-normal modes with spectrum $-i (2 +n) \sqrt{M}$ (where $n = 0,1,2,\cdots$). Asymptotically, all $Q_1^{2+n}(x)$ falls off like $1/x^2$ as $x\rightarrow\infty$ when $n$ is a non-negative integer, and therefore $x_{s,1}$ falls off as $\sigma^{-3}$ as $\sigma\rightarrow\infty$. As mentioned, $x_{s,2}$ also satisfies the same homogeneous equation but we can set it to zero since its amplitudes can be absorbed into those of $x_{s,1}$. Let us choose $A_0 \neq 0$ and $A_i = 0$ for $i\geq 1$ for the sake of illustration. 

In addition to the boundary conditions, we should set an initial condition to get a unique solution. Here we will follow an alternative strategy. In order to get a unique solution at third and higher orders, we will demand that they vanish  faster than $\sigma^{-3}$ as $\sigma\rightarrow \infty$ so that it removes the ambiguity of adding homogeneous pieces which falls off as $\sigma^{-3}$ as noted above. Then the solution for $x_{s,3}$ is  
\begin{eqnarray}\label{Eq:xs3AdS}
   x_{s,3} &=& A_0^3 e^{-6\sqrt{M}\tau} \frac{4(7M - 81 \sigma^2)}{21\sigma^3(\sigma^2 -M)^3}+A_0 M \lambda^2 e^{-2\sqrt{M}\tau} \frac{1}{4\sigma^3(\sigma^2 -M)} \nonumber\\
   && + A_0 M \lambda \gamma_k e^{-3\sqrt{M}\tau} \frac{M -5\sigma^2}{\sigma^3(\sigma^2 -M)^{\frac{3}{2}}} +  A_0 M^2 \gamma_k^2 e^{-4\sqrt{M}\tau} \frac{M -7\sigma^2}{\sigma^3(\sigma^2 -M)^2}.
\end{eqnarray}
The first term above is exactly the first non-linear correction to the solution of the Nambu-Goto equations when we consider the same boundary conditions for the latter as discussed in the next subsection. Note that, the second term proportional to $A_0\lambda^2$ implies that we get further corrections even when the rigid parameter $\gamma_k$ vanishes. Similarly, we get unique solutions for $x_{s,4}$, etc. which reproduce the non-linear corrections to the solutions of the Nambu-Goto equation when $\lambda$ and the rigid parameters vanish. Explicitly, $\tau_{a,2}$ and $\sigma_{a,2}$ are
\begin{eqnarray}\label{Eq:t2x2AdS}
    \tau_{a,2} &=& - A_0\lambda e^{-2\sqrt{M}\tau}\frac{M+\sigma^2}{2M^{3/2}(\sigma^2-M)}- 2A_0\gamma_k e^{-3\sqrt{M}\tau} \frac{\sqrt{M}}{(\sigma^2 -M)^{\frac{3}{2}}},\nonumber\\
    \sigma_{a,2} &=& - A_0\lambda e^{-2\sqrt{M}\tau}\frac{M+\sigma^2}{M\sigma}- 2A_0\gamma_k e^{-3\sqrt{M}\tau}  \frac{M}{\sigma\sqrt{\sigma^2 -M}}, \,\,{\rm etc.}
\end{eqnarray}
Note that
\begin{eqnarray}\label{Eq:t2x2AdSinf}
\tau_{a,2} = -\frac{A_0\lambda}{2{M^{3/2}}}e^{-2\sqrt{M}\tau}+\mathcal{O}\left(\frac{1}{\sigma} \right), \quad \sigma_{a,2} = -\frac{A_0\lambda}{M}e^{-2\sqrt{M}\tau}\sigma +\mathcal{O}\left(\frac{1}{\sigma} \right).
\end{eqnarray}
Therefore, we note that even when we impose the Dirichlet boundary conditions for both $x_1$ and $x_2$ (the coordinates transverse to $\Sigma_1$ and $\Sigma_2$, respectively) so that $\alpha_k = \gamma_k = 0$, there is a \textit{non-trivial relative time-reparametrization at the boundary which encodes the solution of the Nambu-Goto equation corresponding to the bulk junction.} Furthermore, this is true even when we set $\alpha_h$ and $\gamma_h$, which parametrize the linearized isometries, to zero. (In the above case, the time-reparametrization at the boundary reveals that we have turned on only the lowest order quasi-normal mode at the linear order.)

Furthermore, we note that when  $\alpha_{k} =\gamma_{k} = 0 $, we have
\begin{align}\label{Eq:t3x3AdS}
\tau_{a,3} =& \frac{M \gamma_h \left( 48 A_0^2 ~e^{-5\sqrt{M} \tau}+ 3 \lambda^2 ~ e^{-\sqrt{M} \tau} (\sigma^2 -M)^2-4 \gamma_h^2 ~e^{-3\sqrt{M} \tau} \sigma^2 (\sigma^4 -4 M \sigma^2 +3 M^2) \right)}{24 \sigma(\sigma^2 -M)^{5/2}}, \nonumber \\
\sigma_{a,3} =& \frac{M^{3/2} \gamma_h \left( 16 A_0^2 ~e^{-5\sqrt{M} \tau}+  \lambda^2 ~ e^{-\sqrt{M} \tau} (\sigma^2 -M)^2-4 \gamma_h^2 ~e^{-3\sqrt{M} \tau} \sigma^2 (\sigma^4 - M^2) \right)}{8 \sigma^2(\sigma^2 -M)^{3/2}}. 
\end{align}
Near the boundary $\sigma \rightarrow \infty$, these functions behave as
\begin{align}\label{Eq:t3x3AdSinf}
\tau_{a,3} =  -\frac{1}{6} M \gamma_h^3~ e^{-3 \sqrt M \tau} +\mathcal{O}\left(\frac{1}{\sigma} \right), ~~
\sigma_{a,3} = -\frac{1}{2} M^{3/2} \gamma_h^3 ~e^{-3 \sqrt M \tau} \sigma +\mathcal{O}\left(\frac{1}{\sigma} \right). 
\end{align}
Therefore, even when we set $\lambda =0$ and impose the Dirichlet boundary conditions on both $x_1$ and $x_2$, there is a non-trivial time-reparametrization at the boundary at higher orders as well even when the tension vanishes.

\paragraph{Discontinuity of the extrinsic curvature at second and higher orders:} The junction conditions \eqref{Eq:gluing2} giving the discontinuity of the extrinsic curvature at the junction in terms of the worldsheet stress tensor are of the form
\begin{align}
  \ddot{x}_{a,n} - \sigma(\sigma^2 - M)x^\prime_{a,n}= \mathcal{B}^A_{n1},\label{Eq:Kdiscn1}\\
 \dot{x}^{\prime}_{a,n} -\frac{M\dot{x}_{a,n}}{\sigma(\sigma^2 - M)} =\mathcal{B}^A_{n2},\label{Eq:Kdiscn2}\\
    x^{\prime\prime}_{a,n} +\frac{3\sigma^2 -2M}{ \sigma(\sigma^2-M)} x^\prime_{a,n} =\mathcal{B}^A_{n3},\label{Eq:Kdiscn3}
\end{align}
for $n\geq 2$ with $\mathcal{B}^A_{ni}$ being sources constituted by lower order terms and $\lambda$. These equations are in nested form. Solving \eqref{Eq:Kdiscn3} first gives
\begin{equation}\label{Eq:xanAdS}
    x_{a,n} = \tilde{s}_n(\tau,\sigma) + \frac{\sqrt{\sigma^2 -M}}{\sigma}k_{n1}(\tau) + k_{n2}(\tau),
\end{equation}
where $\tilde{s}_n(\tau,\sigma)$ is fully determined by $\mathcal{B}^A_{n3}$, and $k_{n1}$ and $k_{n2}$ are undetermined functions of $\tau$. Substituting \eqref{Eq:xanAdS} into \eqref{Eq:Kdiscn2}, we simply get
\begin{equation}
    \dot{k}_{n2}(\tau) = 0,
\end{equation}
and then substituting \eqref{Eq:xanAdS} into \eqref{Eq:Kdiscn1} and using the above, we get
\begin{equation}
    \ddot{k}_{n1}(\tau) - M {k}_{n1}(\tau) = K_n(\tau)
\end{equation}
with $K_n(\tau)$ determined by lower order terms when $\beta_h = \beta_k = 0$. Both of these are highly non-trivial. In fact the simultaneous solutions of \eqref{Eq:Kdiscn1}, \eqref{Eq:Kdiscn2} and \eqref{Eq:Kdiscn3} exist only because the junction conditions \eqref{Eq:gluing2} are not independent of each other as discussed before and as we have seen previously also in the case of flat space. The general solutions of $k_{n1}$ and $k_{n2}$ are
\begin{align}
&k_{n1}(\tau) = \beta_{k,n} e^{\sqrt{M}\tau} + \gamma_{k,n} e^{-\sqrt{M}\tau} \nonumber\\&
        + \frac{e^{\sqrt{M}\tau} }{2\sqrt{M}}\int_0^\tau {\rm d}\tau_1 \, e^{-\sqrt{M}\tau_1} K_n(\tau_1) -\frac{e^{-\sqrt{M}\tau} }{2\sqrt{M}}\int_0^\tau {\rm d}\tau_1 \, e^{\sqrt{M}\tau_1} K_n(\tau_1),\nonumber\\
        & k_{n2}(\sigma) = \alpha_{k,n}.
\end{align}
Comparing \eqref{Eq:xanAdS} with \eqref{Eq:xa1AdS}, we readily see that $\alpha_{k,n}$, $\beta_{k,n}$ and $\gamma_{k,n}$ can be absorbed into the parameters $\alpha_k$, $\beta_k$ and $\gamma_k$ which give infinitesimal rigid deformations of the hypersurface that preserve its extrinsic curvature. So we can set $\alpha_{k,n}$, $\beta_{k,n}$ and $\gamma_{k,n}$ to zero. We note again that both $\beta_h$ and $\beta_k$ should vanish for \eqref{Eq:Kdiscn1}, \eqref{Eq:Kdiscn2} and \eqref{Eq:Kdiscn3} to have solutions similar to the case of the junction conditions related to the continuity of the induced metric. Thus we can determine $x_{a,n}$ to all orders. As for an illustration, explicitly
\begin{equation}
    x_{a,2} = 4 A_0 \alpha_h e^{-2\sqrt{M}\tau}\frac{M}{\sigma(\sigma^2 - M)}+ 8 A_0\gamma_h e^{-3\sqrt{M}\tau}\frac{M^{\frac{1}{2}}}{(\sigma^2 - M)^{\frac{3}{2}}}, \,\, {\rm etc.}
\end{equation}
for the choice of solution of $x_s$ corresponding to $A_0\neq 0, A_{i\geq 1}=0$ (cf. Eq. \eqref{solxs1}).

The above discussion should make it clear that $x_{a,n}$ for $n\geq 2$ are completely determined by the choice of solution for $x_s$, which is the only degree of freedom, and the four rigid parameters, namely $\alpha_h$, $\gamma_h$, $\alpha_k$ and $\gamma_k$. So, it is not obvious that we can satisfy the Dirichlet boundary conditions for $x_a$ given by \eqref{Eq:DBCxa}. However, just like in the case of $x_{a,1}$, we can satisfy \eqref{Eq:DBCxa} to higher orders simply by requiring that $\alpha_k = \gamma_k = 0$. In this case, we are only left with two rigid parameters, namely $\alpha_h$ and $\gamma_h$ corresponding to worldsheet isometries at the leading order. 

We also observe that the Dirichlet boundary conditions for $\tau_a$ and $\sigma_a$, the relative time and spatial coordinates of the worldsheet, can be imposed only if we choose the trivial solution for $x_s$, namely $x_s = x_0$ (so that $x_{s,i} = 0$ for $i \geq 1$) as should be clear from the asymptotic behavior of $\tau_{a,2}$ and $\sigma_{a,2}$ (see Eq. \eqref{Eq:t2x2AdSinf}), and also set $\alpha_h$ and $\gamma_h$ to zero as should be clear from the asymptotic behavior of $\tau_{a,1}$ and $\sigma_{a,1}$ (see Eq. \eqref{Eq:t1x1AdS}). In this case, we obtain only a unique static solution for $x_a$ which is determined only by the tension and which vanishes when the tension goes to zero. This static solution agrees with the solution reported in \cite{Bachas:2021fqo} where Dirichlet boundary condition was imposed for all four variables, namely $x_s$, $x_a$, $\tau_a$ and $\sigma_a$. 

As discussed before, the Dirichlet boundary conditions on $\tau_a$ and $\sigma_a$ can be relaxed because the Dirichlet boundary conditions on $x_s$ and $x_a$ are enough to ensure that both $\Sigma_1$ and $\Sigma_2$ end at a common spatial point at the boundary of the full spacetime although the time coordinates at the boundaries of these hypersurfaces can be non-trivially related to each other. Therefore only two rigid parameters, namely $\alpha_h$ and $\gamma_h$, related to the worldsheet isometries at the leading order, parametrize such general solutions with Dirichlet boundary conditions on both $\Sigma_1$ and $\Sigma_2$. In fact, as discussed above (recall the discussion about $\tau_{a,2}$), \textit{the boundary time-reparametrization persists to higher orders even when we set $\alpha_h$ and $\gamma_h$ to zero, and the solution of the Nambu-Goto solution corresponding to the bulk junction can actually be decoded from the relative time-reparametrization at the boundary.}

We note from the explicit expressions above that the perturbation expansion breaks down near the worldsheet horizon $\sigma = \sqrt{M}$. The same is true for the solutions of the Nambu-Goto equation.

\subsection{Matching with the solutions of the Nambu-Goto equation}
To compare with the solutions of the Nambu-Goto equation, we proceed as in the flat space case by choosing the following parametric form of a hypersurface:
\begin{equation}
    t = \tau, \quad z = \sigma, \quad x = f(\tau, \sigma).
\end{equation}
The Nambu-Goto equation for this hypersurface in the background metric \eqref{Eq:BTZm} is 
\begin{align}\label{Eq:NGFS}
  &\left(\frac{1}{\sigma^2 -M} +\sigma^2 {f^\prime}^2\right) \ddot{f} -  \left(\sigma^2 -M -\sigma^2 \dot{f}^2\right) f^{\prime\prime} - \frac{2(2\sigma^2-M)}{\sigma}f^\prime \nonumber\\&
  - 2 \sigma^2 \dot{f}f^\prime \dot{f}^\prime + \sigma\left(4+\frac{3M}{\sigma^2 - M}\right)\dot{f}^2f^\prime - \sigma(M^2 - 3M\sigma^2+ 2\sigma^4){f^\prime}^3= 0.
\end{align}
With the perturbative expansion for $f$ given by \eqref{Eq:fpert}, we obtain the following equations:
\begin{eqnarray}\label{Eq:feqnpert}
    \frac{1}{\sigma^2 -M} \ddot{f}_{n} - (\sigma^2 -M){f}^{\prime\prime}_n -\frac{2(2\sigma^2 -M)}{\sigma}f^\prime_n = {\mathcal{S}}^A_n
\end{eqnarray}
where the sources ${\mathcal{S}}^A_n$ vanish for $n= 1$ and is constituted of lower order terms for $n\geq 3$. Since the Nambu-Goto equation is odd in $f$, we can set $f_{n}=0$ for positive even integral values of $n$. Also, ${\mathcal{S}}^A_n$ vanishes for all positive integral values of $n$. For odd values of $n$, we can solve the above equations by demanding Dirichlet boundary conditions for $f_n$, i.e. they vanish at $\sigma =\infty$, and that they satisfy the ingoing boundary condition at the worldsheet horizon as discussed in the context of $x_{s,n}$. Furthermore, in order to obtain $f_{n}$ uniquely for $n\geq 3$, we need to further demand that it vanishes faster than $\sigma^{-3}$ as $\sigma\rightarrow\infty$.

When $\lambda$ and the rigid parameters $\alpha_h$, $\gamma_h$, $\alpha_k$ and $\gamma_k$ vanish, $x_{s,n}$ also vanishes for positive even integral values of $n$ like $f_n$, and for positive odd integral values, the equations of $x_{s,n}$ given by \eqref{Eq:xsAdSn} coincide with that those of $f_n$ given by \eqref{Eq:feqnpert}. Therefore, \textit{$x_s$ coincides exactly with $f$ when $\lambda$ and the rigid parameters $\alpha_h$, $\gamma_h$, $\alpha_k$ and $\gamma_k$ vanish.} As for illustration, if we keep only the lowest quasi-normal mode for $f_1$, then $f_3$ is just the first term of $x_{s,3}$ given in \eqref{Eq:xs3AdS} which is the only surviving term when $\lambda$ and the rigid parameters $\alpha_h$, $\gamma_h$, $\alpha_k$ and $\gamma_k$ vanish. 

Since $x_{a}$, $\tau_a$ and $\sigma_a$ are determined by the chosen solution of $x_s$ and the rigid parameters as demonstrated earlier, it follows that there is a one-to-one correspondence between the solutions of the junction conditions and the Nambu-Goto equations. We have checked that this correspondence holds up to the seventh order in the perturbative expansion. As discussed before, the number of rigid parameters is only two, namely $\alpha_h$ and $\gamma_h$ if we further impose Dirichlet boundary conditions on both the hypersurfaces. 

 \AM{It is not clear whether the limit $\lambda\rightarrow 0$ produces a smooth spacetime in the presence of rigid deformations} since the relative transverse coordinate $x_a$ does not vanish \AM{in this case when the tensionless limit is taken}. These solutions are still in one-to-one correspondence with the solutions of Nambu-Goto equations up to the rigid deformations.

We also want to emphasize that the tensionless limit produces non-trivial solutions even when we impose the Dirichlet boundary condition identically on both the hypersurfaces so that $\alpha_k = \gamma_k =0$ and we have only the deformation parameters $\alpha_h$ and $\gamma_h$ related to the worldsheet isometries at the leading order. Nevertheless, when all the rigid parameters vanish implying that the Dirichlet boundary condition is imposed on all the four variables ($\tau_a$, $\sigma_a$, $x_s$ and $x_a$), the solutions of the junction conditions are trivial in the tensionless limit as $x_a$, $\tau_a$ and $\sigma_a$ vanish leading to a smooth spacetime, while $x_s$ coincides \textit{exactly} with a solution of the Nambu-Goto equation representing just a probe string in this spacetime.\footnote{However, when $\lambda \neq 0$, the Dirichlet boundary condition can be imposed on $\tau_{a,n}$ and $\sigma_{a,n}$ (recall the case of $\tau_{a,2}$) only if we choose the trivial solution for $x_s$, which is $x_s = x_0$, implying that all the fluctuations corresponding to the Nambu-Goto quasi-normal modes should vanish. For non-vanishing $\lambda$, we simply recover the known static solution for $x_a$, while $\tau_a$ and $\sigma_a$ vanish as discussed before, when all the four variables satisfy the Dirichlet boundary condition.}

\section{Discussion}\label{Sec:Disc}

Using a perturbative approach in which the string tension and the amplitudes of fluctuations of the hypersurfaces from a common configuration with vanishing extrinsic curvature are small, we have demonstrated that the Nambu-Goto equation directly emerges from the junction conditions of gravitational equations both in a locally AdS and a locally flat spacetime with three spacetime dimensions. Here, we have glued two identical copies of a spacetime. However, there are other possible solutions in which two different Ba{\~n}ados spacetimes \cite{Banados:1998gg}, e.g. two BTZ black holes with different masses and angular momentum can be glued. Solutions of such type have been studied in \cite{Bachas:2020yxv,Bachas:2021fqo,Bachas:2021tnp}, and solutions with null interfaces have been explored in \cite{Kibe:2021qjy,Banerjee:2022dgv}. We should also study solutions where the two spacetimes glued at the junction can have different cosmological constants. 

 While we have obtained our results by solving the perturbative expansion up to eighth order, it would be very interesting to have a non-perturbative proof of our statements. This would probably uncover some deep structural reasons underlying the correspondence between the Nambu-Goto equations and the junction conditions.

A natural question is the interpretation of our solutions and their generalizations within the AdS$_3$/CFT$_2$ correspondence. We recall that when we glue two copies of a BTZ black hole and set Dirichlet boundary conditions for all the four variables $\tau_a$, $\sigma_a$, $x_s$ and $x_a$, we obtain a unique static solution in which $x_s$ is a trivial constant solution of the Nambu-Goto equation while $\tau_a$ and $\sigma_a$ vanish, and $x_a$ is determined just by the tension. This solution corresponds to a defect \cite{Wong:1994np,Oshikawa:1996dj,Billo:2016cpy} in the dual conformal field theory with large central charge, where the string tension $\lambda$ is simply a parameter that characterizes the defect. The tension determines the full spacetime, and thus the dual correlation functions. Furthermore, it was shown in \cite{Bachas:2021tnp} that the bulk junction conditions indeed reproduce the reflection and transmission coefficients of the defect which had been analyzed in \cite{Quella:2006de,Kimura:2015nka,Meineri:2019ycm}.

Our general solutions which correspond to the Nambu-Goto solutions up to the rigid deformations should be interpreted as general state-dependent interfaces in the dual conformal field theory. Note that the Nambu-Goto equation has no intrinsic dimensionful parameters, and the quasinormal mode spectrum that characterize the solutions is determined by the mass of the background spacetime, i.e. the dual state. When we impose the Dirichlet boundary condition identically on both the hypersurfaces (i.e. for both $x_s$ and $x_a$), the solutions should holographically represent dynamical interfaces between two copies of the same state in the dual CFT, particularly two thermal states with the same temperature since the BTZ black hole is dual to a thermal state. No transfer of energy and momentum occur through the interfaces. Nevertheless, the interfaces have their own dynamics, which are characterized by the solutions of the Nambu-Goto equations and the rigid parameters of the bulk solution. All these parameters and the solutions of the Nambu-Goto equation together with the tension determine the correlation functions of the full system. In this aspect, it could be important to recall that two sides of the interface necessarily have a relative time-reparametrization (given by the asymptotic limit of $\tau_{a}$) which can decode the non-trivial Nambu-Goto solution in the bulk. This feature is not present in a usual defect. In this context, it is important to understand such interfaces in the vacuum which correspond to bulk junctions in pure AdS$_3$.

{The holographic interpretation of our solutions would be important to understand whether the rigid parameters are related to the physical properties of the interface. The appropriate boundary conditions for the junction corresponding to the dual interface set the values of some of the rigid physical parameters (as we have seen in the context of the Dirichlet boundary condition in AdS). The remaining rigid parameters may actually characterize physical properties of the interface. Although the latter can be absorbed via relative isometries at the leading order, they might correspond to physically inequivalent solutions (as in the case of improper diffeomorphisms in AdS/CFT). In the context of AdS, these issues can also be addressed by the computation of holographic entanglement and Renyi entropies, which can directly reveal the full bi-partite state in the CFT glued at the interface, and also clarify whether the tensionless limit leads to a trivial interface.}

In the future, we would like to construct such interfaces between two identical thermal states in two-dimensional conformal field theories and generalizations of such setups because such constructions can reveal how the fundamental string of the dual string theory emerges directly from the conformal field theory even beyond the strong coupling and large central charge limit. As a special case, it would be also interesting to see how the solutions of the junction conditions discussed here which appear in the limit of vanishing string tension correspond to non-trivial dynamical interfaces in the strongly coupled conformal field theory with large central charge.

The general solutions describing junctions between two different Ba{\~n}ados spacetimes with the same or different (negative) cosmological constants should represent interfaces between two different states in the same or different CFTs and these could give fresh insights into bulk reconstruction in holographic duality (see \cite{Harlow:2018fse,Kibe:2021gtw} for reviews). Furthermore, these constructions should have their own utilities in the study of quantum thermodynamics of non-equilibrium ensembles and quantum engines. In the latter aspect, the study of null junctions has shown that holography can give novel understanding of quantum thermodynamics of irreversible entropy production in phase transitions in many-body systems and also in quantum channels like the Landauer erasure implemented in many-body systems \cite{Kibe:2021qjy,Banerjee:2022dgv} (in the latter case it has shown the utility of novel non-isometric dense encodings for quantum memories which cannot be erased in microscopic time-scales).

{A related question is whether we can go beyond a perturbative approach and formulate a way to solve the gravitational equations with the junction conditions via numerical relativity setting necessary initial conditions and also boundary conditions. In fact, a fundamental question in this context is which types of initial data would lead to non-singular spacetimes such as those not having closed time-like curves. Such an initial value formulation should lead to the most general constructions of gravitational solutions involving junctions which can be interpreted as novel interfaces holographically.}

Another fundamental question is whether we can quantize our three-dimensional gravitational solutions with junctions generalizing the approach in \cite{Kim:2015qoa,Cotler:2018zff,Collier:2023fwi} \AM{(based on rewriting three-dimensional gravity in AdS as PSL$(2,\mathbb{R})$ $\times$ PSL$(2,\mathbb{R})$ Chern-Simons theory \cite{Achucarro:1986uwr,Witten:1988hc} and further work in \cite{Krasnov:2005dm,Scarinci:2011np})} and obtain part of the spectrum of first quantized bosonic string theory in background flat space \cite{Polchinski:1998rq,Polchinski:1998rr} or AdS space \cite{Maldacena:2000hw,Maldacena:2000kv,Gaberdiel:2018rqv} within the full Hilbert space. The latter should however also include boundary gravitons that scatter off the quantum string. Aspects of quantum string theory can thus emerge directly from lower dimensional gravity together with additional degrees of freedom. The presence of rigid deformations of the Nambu-Goto equations arising from the junction conditions implies that the quantized string theory obtained from quantizing such gravitational solutions can have other non-trivial features as well. Of course, we can generalize the junctions in the context of gauged supergravities and wonder if some aspects of classical and quantum superstring theory can emerge similarly from super-spacetime dynamics in three dimensions.

\begin{acknowledgments}
We would like to thank Costas Bachas, Jewel Ghosh, Rajesh Gopakumar, \AB{and Marc Henneaux for insightful discussions and comments on the manuscript. We also thank Arnab Kundu for his collaboration during the initial stages of the project.} The research of AB is implemented in the framework of H.F.R.I. call “Basic research Financing (Horizontal support of all Sciences)” under the National Recovery and Resilience Plan “Greece 2.0” funded by the European Union – NextGenerationEU (H.F.R.I. Project Number: 15384). AM acknowledges support from Fondecyt grant 1240955.
\end{acknowledgments}

\bibliographystyle{JHEP} 
 \bibliography{biblio}

\providecommand{\href}[2]{#2}\begingroup\raggedright\begin{thebibliography}{10}

\bibitem{Polchinski:1998rq}
J.~Polchinski, \emph{{String theory. Vol. 1: An introduction to the bosonic
  string}}, Cambridge Monographs on Mathematical Physics, Cambridge University
  Press (12, 2007),
  \href{https://doi.org/10.1017/CBO9780511816079}{10.1017/CBO9780511816079}.

\bibitem{Polchinski:1998rr}
J.~Polchinski, \emph{{String theory. Vol. 2: Superstring theory and beyond}},
  Cambridge Monographs on Mathematical Physics, Cambridge University Press (12,
  2007),
  \href{https://doi.org/10.1017/CBO9780511618123}{10.1017/CBO9780511618123}.

\bibitem{Maldacena:2000hw}
J.M.~Maldacena and H.~Ooguri, \emph{{Strings in AdS(3) and SL(2,R) WZW model
  1.: The Spectrum}}, \href{https://doi.org/10.1063/1.1377273}{\emph{J. Math.
  Phys.} {\bfseries 42} (2001) 2929}
  [\href{https://arxiv.org/abs/hep-th/0001053}{{\ttfamily hep-th/0001053}}].

\bibitem{Maldacena:2000kv}
J.M.~Maldacena, H.~Ooguri and J.~Son, \emph{{Strings in AdS(3) and the SL(2,R)
  WZW model. Part 2. Euclidean black hole}},
  \href{https://doi.org/10.1063/1.1377039}{\emph{J. Math. Phys.} {\bfseries 42}
  (2001) 2961} [\href{https://arxiv.org/abs/hep-th/0005183}{{\ttfamily
  hep-th/0005183}}].

\bibitem{Gaberdiel:2018rqv}
M.R.~Gaberdiel and R.~Gopakumar, \emph{{Tensionless string spectra on
  AdS$_{3}$}}, \href{https://doi.org/10.1007/JHEP05(2018)085}{\emph{JHEP}
  {\bfseries 05} (2018) 085}
  [\href{https://arxiv.org/abs/1803.04423}{{\ttfamily 1803.04423}}].

\bibitem{Kraus:2006wn}
P.~Kraus, \emph{{Lectures on black holes and the AdS(3) / CFT(2)
  correspondence}}, {\emph{Lect. Notes Phys.} {\bfseries 755} (2008) 193}
  [\href{https://arxiv.org/abs/hep-th/0609074}{{\ttfamily hep-th/0609074}}].

\bibitem{Eberhardt:2018ouy}
L.~Eberhardt, M.R.~Gaberdiel and R.~Gopakumar, \emph{{The Worldsheet Dual of
  the Symmetric Product CFT}},
  \href{https://doi.org/10.1007/JHEP04(2019)103}{\emph{JHEP} {\bfseries 04}
  (2019) 103} [\href{https://arxiv.org/abs/1812.01007}{{\ttfamily
  1812.01007}}].

\bibitem{Eberhardt:2019ywk}
L.~Eberhardt, M.R.~Gaberdiel and R.~Gopakumar, \emph{{Deriving the
  AdS$_{3}$/CFT$_{2}$ correspondence}},
  \href{https://doi.org/10.1007/JHEP02(2020)136}{\emph{JHEP} {\bfseries 02}
  (2020) 136} [\href{https://arxiv.org/abs/1911.00378}{{\ttfamily
  1911.00378}}].

\bibitem{Karch:2000ct}
A.~Karch and L.~Randall, \emph{{Locally localized gravity}},
  \href{https://doi.org/10.1088/1126-6708/2001/05/008}{\emph{JHEP} {\bfseries
  05} (2001) 008} [\href{https://arxiv.org/abs/hep-th/0011156}{{\ttfamily
  hep-th/0011156}}].

\bibitem{DeWolfe:2001pq}
O.~DeWolfe, D.Z.~Freedman and H.~Ooguri, \emph{{Holography and defect conformal
  field theories}},
  \href{https://doi.org/10.1103/PhysRevD.66.025009}{\emph{Phys. Rev. D}
  {\bfseries 66} (2002) 025009}
  [\href{https://arxiv.org/abs/hep-th/0111135}{{\ttfamily hep-th/0111135}}].

\bibitem{Takayanagi:2011zk}
T.~Takayanagi, \emph{{Holographic Dual of BCFT}},
  \href{https://doi.org/10.1103/PhysRevLett.107.101602}{\emph{Phys. Rev. Lett.}
  {\bfseries 107} (2011) 101602}
  [\href{https://arxiv.org/abs/1105.5165}{{\ttfamily 1105.5165}}].

\bibitem{Bachas:2020yxv}
C.~Bachas, S.~Chapman, D.~Ge and G.~Policastro, \emph{{Energy Reflection and
  Transmission at 2D Holographic Interfaces}},
  \href{https://doi.org/10.1103/PhysRevLett.125.231602}{\emph{Phys. Rev. Lett.}
  {\bfseries 125} (2020) 231602}
  [\href{https://arxiv.org/abs/2006.11333}{{\ttfamily 2006.11333}}].

\bibitem{Bachas:2021tnp}
C.~Bachas, Z.~Chen and V.~Papadopoulos, \emph{{Steady states of holographic
  interfaces}}, \href{https://doi.org/10.1007/JHEP11(2021)095}{\emph{JHEP}
  {\bfseries 11} (2021) 095}
  [\href{https://arxiv.org/abs/2107.00965}{{\ttfamily 2107.00965}}].

\bibitem{Liu:2024oxg}
Y.~Liu, H.-D.~Lyu and C.-Y.~Wang, \emph{{On AdS$_3$/ICFT$_2$ with a dynamical
  scalar field located on the brane}},
  \href{https://arxiv.org/abs/2403.20102}{{\ttfamily 2403.20102}}.

\bibitem{Israel:1966rt}
W.~Israel, \emph{{Singular hypersurfaces and thin shells in general
  relativity}}, \href{https://doi.org/10.1007/BF02710419}{\emph{Nuovo Cim. B}
  {\bfseries 44S10} (1966) 1}.

\bibitem{Brown:1992br}
J.D.~Brown and J.W.~York, Jr., \emph{{Quasilocal energy and conserved charges
  derived from the gravitational action}},
  \href{https://doi.org/10.1103/PhysRevD.47.1407}{\emph{Phys. Rev. D}
  {\bfseries 47} (1993) 1407}
  [\href{https://arxiv.org/abs/gr-qc/9209012}{{\ttfamily gr-qc/9209012}}].

\bibitem{Bachas:2021fqo}
C.~Bachas and V.~Papadopoulos, \emph{{Phases of Holographic Interfaces}},
  \href{https://doi.org/10.1007/JHEP04(2021)262}{\emph{JHEP} {\bfseries 04}
  (2021) 262} [\href{https://arxiv.org/abs/2101.12529}{{\ttfamily
  2101.12529}}].

\bibitem{Banados:1992wn}
M.~Banados, C.~Teitelboim and J.~Zanelli, \emph{{The Black hole in
  three-dimensional space-time}},
  \href{https://doi.org/10.1103/PhysRevLett.69.1849}{\emph{Phys. Rev. Lett.}
  {\bfseries 69} (1992) 1849}
  [\href{https://arxiv.org/abs/hep-th/9204099}{{\ttfamily hep-th/9204099}}].

\bibitem{Banados:1992gq}
M.~Banados, M.~Henneaux, C.~Teitelboim and J.~Zanelli, \emph{{Geometry of the
  (2+1) black hole}},
  \href{https://doi.org/10.1103/PhysRevD.48.1506}{\emph{Phys. Rev. D}
  {\bfseries 48} (1993) 1506}
  [\href{https://arxiv.org/abs/gr-qc/9302012}{{\ttfamily gr-qc/9302012}}].

\bibitem{Banados:1998gg}
M.~Banados, \emph{{Three-dimensional quantum geometry and black holes}},
  \href{https://doi.org/10.1063/1.59661}{\emph{AIP Conf. Proc.} {\bfseries 484}
  (1999) 147} [\href{https://arxiv.org/abs/hep-th/9901148}{{\ttfamily
  hep-th/9901148}}].

\bibitem{Kibe:2021qjy}
T.~Kibe, A.~Mukhopadhyay and P.~Roy, \emph{{Quantum Thermodynamics of
  Holographic Quenches and Bounds on the Growth of Entanglement from the
  Quantum Null Energy Condition}},
  \href{https://doi.org/10.1103/PhysRevLett.128.191602}{\emph{Phys. Rev. Lett.}
  {\bfseries 128} (2022) 191602}
  [\href{https://arxiv.org/abs/2109.09914}{{\ttfamily 2109.09914}}].

\bibitem{Banerjee:2022dgv}
A.~Banerjee, T.~Kibe, N.~Mittal, A.~Mukhopadhyay and P.~Roy, \emph{{Erasure
  Tolerant Quantum Memory and the Quantum Null Energy Condition in Holographic
  Systems}}, \href{https://doi.org/10.1103/PhysRevLett.129.191601}{\emph{Phys.
  Rev. Lett.} {\bfseries 129} (2022) 191601}
  [\href{https://arxiv.org/abs/2202.00022}{{\ttfamily 2202.00022}}].

\bibitem{Wong:1994np}
E.~Wong and I.~Affleck, \emph{{Tunneling in quantum wires: A Boundary conformal
  field theory approach}},
  \href{https://doi.org/10.1016/0550-3213(94)90479-0}{\emph{Nucl. Phys. B}
  {\bfseries 417} (1994) 403}
  [\href{https://arxiv.org/abs/cond-mat/9311040}{{\ttfamily
  cond-mat/9311040}}].

\bibitem{Oshikawa:1996dj}
M.~Oshikawa and I.~Affleck, \emph{{Boundary conformal field theory approach to
  the critical two-dimensional Ising model with a defect line}},
  \href{https://doi.org/10.1016/S0550-3213(97)00219-8}{\emph{Nucl. Phys. B}
  {\bfseries 495} (1997) 533}
  [\href{https://arxiv.org/abs/cond-mat/9612187}{{\ttfamily
  cond-mat/9612187}}].

\bibitem{Billo:2016cpy}
M.~Bill\`o, V.~Gon\c{c}alves, E.~Lauria and M.~Meineri, \emph{{Defects in
  conformal field theory}},
  \href{https://doi.org/10.1007/JHEP04(2016)091}{\emph{JHEP} {\bfseries 04}
  (2016) 091} [\href{https://arxiv.org/abs/1601.02883}{{\ttfamily
  1601.02883}}].

\bibitem{Quella:2006de}
T.~Quella, I.~Runkel and G.M.T.~Watts, \emph{{Reflection and transmission for
  conformal defects}},
  \href{https://doi.org/10.1088/1126-6708/2007/04/095}{\emph{JHEP} {\bfseries
  04} (2007) 095} [\href{https://arxiv.org/abs/hep-th/0611296}{{\ttfamily
  hep-th/0611296}}].

\bibitem{Kimura:2015nka}
T.~Kimura and M.~Murata, \emph{{Transport Process in Multi-Junctions of Quantum
  Systems}}, \href{https://doi.org/10.1007/JHEP07(2015)072}{\emph{JHEP}
  {\bfseries 07} (2015) 072}
  [\href{https://arxiv.org/abs/1505.05275}{{\ttfamily 1505.05275}}].

\bibitem{Meineri:2019ycm}
M.~Meineri, J.~Penedones and A.~Rousset, \emph{{Colliders and conformal
  interfaces}}, \href{https://doi.org/10.1007/JHEP02(2020)138}{\emph{JHEP}
  {\bfseries 02} (2020) 138}
  [\href{https://arxiv.org/abs/1904.10974}{{\ttfamily 1904.10974}}].

\bibitem{Harlow:2018fse}
D.~Harlow, \emph{{TASI Lectures on the Emergence of Bulk Physics in AdS/CFT}},
  \href{https://doi.org/10.22323/1.305.0002}{\emph{PoS} {\bfseries TASI2017}
  (2018) 002} [\href{https://arxiv.org/abs/1802.01040}{{\ttfamily
  1802.01040}}].

\bibitem{Kibe:2021gtw}
T.~Kibe, P.~Mandayam and A.~Mukhopadhyay, \emph{{Holographic spacetime, black
  holes and quantum error correcting codes: a review}},
  \href{https://doi.org/10.1140/epjc/s10052-022-10382-1}{\emph{Eur. Phys. J. C}
  {\bfseries 82} (2022) 463}
  [\href{https://arxiv.org/abs/2110.14669}{{\ttfamily 2110.14669}}].

\bibitem{Kim:2015qoa}
J.~Kim and M.~Porrati, \emph{{On a Canonical Quantization of 3D Anti de Sitter
  Pure Gravity}}, \href{https://doi.org/10.1007/JHEP10(2015)096}{\emph{JHEP}
  {\bfseries 10} (2015) 096}
  [\href{https://arxiv.org/abs/1508.03638}{{\ttfamily 1508.03638}}].

\bibitem{Cotler:2018zff}
J.~Cotler and K.~Jensen, \emph{{A theory of reparameterizations for AdS$_3$
  gravity}}, \href{https://doi.org/10.1007/JHEP02(2019)079}{\emph{JHEP}
  {\bfseries 02} (2019) 079}
  [\href{https://arxiv.org/abs/1808.03263}{{\ttfamily 1808.03263}}].

\bibitem{Collier:2023fwi}
S.~Collier, L.~Eberhardt and M.~Zhang, \emph{{Solving 3d gravity with Virasoro
  TQFT}}, \href{https://doi.org/10.21468/SciPostPhys.15.4.151}{\emph{SciPost
  Phys.} {\bfseries 15} (2023) 151}
  [\href{https://arxiv.org/abs/2304.13650}{{\ttfamily 2304.13650}}].

\bibitem{Achucarro:1986uwr}
A.~Achucarro and P.K.~Townsend, \emph{{A Chern-Simons Action for
  Three-Dimensional anti-De Sitter Supergravity Theories}},
  \href{https://doi.org/10.1016/0370-2693(86)90140-1}{\emph{Phys. Lett. B}
  {\bfseries 180} (1986) 89}.

\bibitem{Witten:1988hc}
E.~Witten, \emph{{(2+1)-Dimensional Gravity as an Exactly Soluble System}},
  \href{https://doi.org/10.1016/0550-3213(88)90143-5}{\emph{Nucl. Phys. B}
  {\bfseries 311} (1988) 46}.

\bibitem{Krasnov:2005dm}
K.~Krasnov and J.-M.~Schlenker, \emph{{Minimal surfaces and particles in
  3-manifolds}}, \href{https://doi.org/10.1007/s10711-007-9132-1}{\emph{Geom.
  Dedicata} {\bfseries 126} (2007) 187}
  [\href{https://arxiv.org/abs/math/0511441}{{\ttfamily math/0511441}}].

\bibitem{Scarinci:2011np}
C.~Scarinci and K.~Krasnov, \emph{{The universal phase space of $AdS_3$
  gravity}}, \href{https://doi.org/10.1007/s00220-012-1655-0}{\emph{Commun.
  Math. Phys.} {\bfseries 322} (2013) 167}
  [\href{https://arxiv.org/abs/1111.6507}{{\ttfamily 1111.6507}}].

\end{thebibliography}\endgroup

\end{document}